\documentclass[aps,pra,floats,floatfix,superscriptaddress,twocolumn]{revtex4-1}
\usepackage{ifthen}
\usepackage{ifpdf}
\usepackage{color}
\ifpdf
\usepackage{graphicx}
\usepackage{epstopdf}
\else
\usepackage{graphicx}
\usepackage{epsfig}
\fi
\usepackage{ulem}
\usepackage{latexsym}
\usepackage{amsmath}
\usepackage{amssymb}
\usepackage{bm}
\usepackage{wasysym}

\usepackage{latexsym}
\usepackage{amsmath}
\usepackage{amssymb}
\usepackage{bm}
\usepackage{wasysym}
\usepackage{ulem}
\usepackage{braket}
\usepackage{hyperref}

\begin{document}
 \title{Correlating two qubits via common cavity environment}

\author{Amit Dey}
\email{amit.dey.85@gmail.com }
\address{Ramananda College, Bankura University, Bankura 722122, India}

	\date{\today}

\begin{abstract}
Generation of quantum entanglement between a pair of qubits is studied in a cavity-QED platform. The qubit pair is placed inside a common cavity environment. We show that the relative strength of qubit-photon couplings is crucial for establishing inter-qubit entanglement. Resonance as well as off-resonance between the qubits and photon are considered . For off-resonant case we detect a threshold value of coupling ratio, beyond which maximally entangled state is always available. The resonant case displays interesting non-monotonic behavior, where the maximum entanglement peaks at an intermediate coupling ratio. The driven-dissipative dynamics of our model exhibits non-trivial dependence of steady-state entanglement on the drive strength.   
\end{abstract}

\maketitle

%\maketitle

\section{Introduction }
Quantum correlations of multi-qubit systems are essential ingredients for quantum technology \cite{Nielsen2010-zt}. Quantum entanglement is one of them and has been extensively exploited in quantum information processing \cite{RevModPhys.81.865}, quantum cryptography \cite{PhysRevLett.67.661}, quantum teleportation \cite{PhysRevLett.70.1895} etc. Entanglement generation is also an avenue to harness nonlocality and its application in quantum information science \cite{PhysicsPhysiqueFizika.1.195,Greenberger1990,PhysRevLett.82.1345,Pan2000}. Preparation of quantum entangled state and its control are vital aspects to access quantum mechanical functionalities \cite{PhysRevLett.98.070502,Bugu2020}.

Cavity QED is a robust platform delivering myriad of applications suitable for quantum technology \cite{RevModPhys.93.025005}. Entangled pair of qubits is well realized in cavity QED \cite{PhysRevLett.79.1,PhysRevA.96.052311,PhysRevA.90.052315, PhysRevLett.92.117902, Rauschenbeutel2000, PhysRevLett.85.2392}. Quantum gate operations have also been demonstrated in cavity-QED-based systems \cite{PhysRevLett.75.4710,PhysRevLett.74.4083}. Various protocols for dissipative preparation and protection of Bell state have been proposed in similar systems \cite{PhysRevA.88.023849,PhysRevLett.106.090502,PhysRevA.84.064302,PhysRevA.88.032317}. In many quantum information processing schemes the cavity works as a memory preserving unit for the cavity-QED system \cite{PhysRevLett.79.769}, where the cavity dissipation usually hampers the quantum mechanical operations of the system. Schemes, where the cavity is only virtually excited, have been proposed to bypass such issues. Such scheme not only establishes inter-qubit quantum correlation without exchanging energy between the qubit and the cavity\cite{PhysRevLett.87.230404,PhysRevLett.85.2392,Zhang2014}, but also can be exploited to design long-range inter-qubit interactions in quantum networks with varied geometry \cite{majer2007,PhysRevA.75.032329,Borjans2019,Ritter2012,Ray2022}. Establishing various degrees of connectivity in an extended lattice is pivotal for state transfer and entanglement generation over large distances. Cavity-mediated interaction among qubits is a key to many such applications. Scalable cavity-QED structures with nontrivial connectivity have been employed to demonstrate distributed quantum informative operations and localization phenomenon \cite{Song2019,Xu2020,Ray2022}. Artificially engineered lattices exhibits fascinating behaviors that can be exploited in efficient quantum simulations\cite{Kollr2019}.

Multi-qubit entanglement generation has been mostly studied for identical cavity-qubit couplings \cite{PhysRevLett.87.230404,PhysRevLett.85.2392,Zhang2014}. However, in many-qubit architecture qubits can be non-identical for various reasons. Some of them arise due to practical experimental challenges, whereas, some are intended to provide additional advantages. Single atom qubits of same species can have varied dipolar transition due to fluctuating charge distribution. Moreover, the spatial separation of atoms can result in varied electromagnetic field they encounter locally \cite{PhysRevA.96.042714}. In circuit-QED systems coupling between qubits and quantized excitation is established by various circuit components such as capacitors, inductors, Josephson junction etc. Fluctuation in these components can lead to considerable change in the the resulting coupling constant \cite{RevModPhys.93.025005}. Furthermore, dual-species qubit systems have been emerged as significant solution to the challenge of qubit-specific read out leading to non-demolition measurement \cite{PhysRevX.12.011040,PhysRevLett.128.083202,https://doi.org/10.48550/arxiv.2401.10325}. It also establishes precise control over individual qubits with negligeable cross talks among the qubits. The impact of qubit asymmetry on entanglement decay rates has been studied and such effect is shown to be dependent on preparation of initial state \cite{Abdelmagid2023}. Driven-dissipative preparation and retention of Bell state has been proposed for non-identical qubit pairs \cite{PhysRevA.88.023849}. These entail investigation of non-identical light-matter interactions in a cavity-QED system that serves the purpose of entanglement generation. Such a study also gives the idea of robustness that the system output exhibits against coupling fluctuations.

In this manuscript we consider a couple of qubits embedded in the same cavity environment with varying degrees of light-matter coupling asymmetry. The objective of the study is to efficiently entangle the qubit pair without direct inter-qubit interaction and scrutinize the effects of asymmetry on resulting entanglement. We also study the behavior of quantum coherence of the qubit system. We consider both the resonant and dispersive cases and seek the favourable conditions for maximizing entanglement. We mainly deal with closed system dynamics for dispersive case where the cavity dissipation remains more or less inactive. For resonant case we consider both the closed and open-system scenarios. The manuscript is presented in the following manner.

In Sec. II we introduce the model and calculate the eigenvalues and eigenvectors of the model Hamiltonian. In Sec. III we present the numerical and analytical results for dispersive (Sec. III A) and resonant (Sec. III B) cases. In Sec. IV open-system dynamics is dealt with. Finally, in Sec. V we conclude our result and discuss future directions.
\section{Model}

\begin{figure}[htb]
    \centering
    \includegraphics[scale=.4]{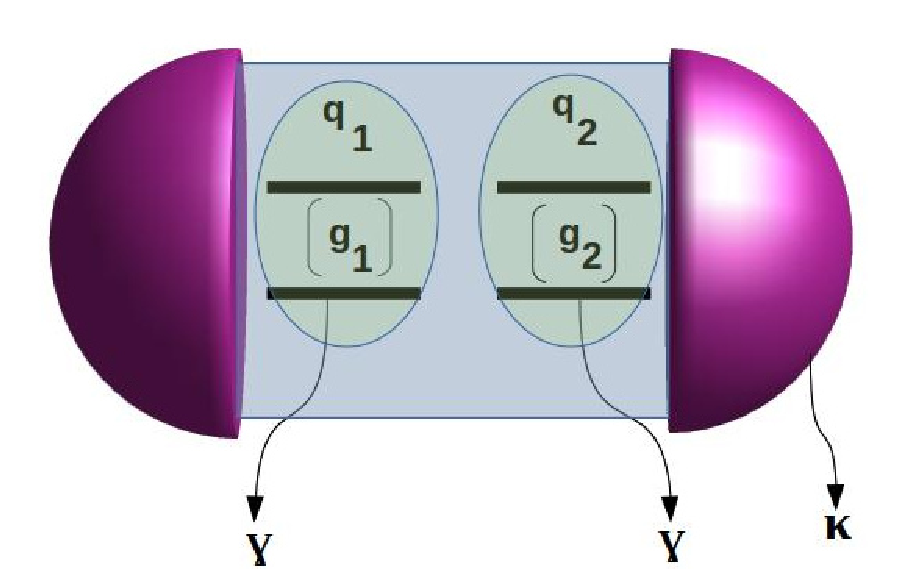}
  %  \captionsetup{justification = centerlast}
    \caption{Schematic diagram depicting two qubits $q_1$ and $q_2$ embedded within a cavity enclosed by the magenta hemispheres. $\gamma$ and $\kappa$ are decay rates for qubit and cavity excitations, respectively. $g_1$ and $g_2$ are the cavity-qubit couplings for the two qubits.}
    \label{schematic}
\end{figure}

The Hamiltonian for our system is given by 
\begin{eqnarray}
 H&=&\omega a^{\dagger}a+\sum_{i=1}^{2} \epsilon_i S_i+\sum_{i=1}^2 g_i (a^{\dagger} S_i^{-}+{\rm h. c.}),
 \label{ham}
\end{eqnarray}
where $a^{\dagger}$ and $S^{+}_i$ are the raising operator for the cavity and qubit degrees of freedom, respectively. $\omega$, $\epsilon_i$, and $g_i$ are photon energy, excitation of i-th qubit, and qubit-photon coupling, respectively. Here, there is no direct interaction between the qubits and the qubits are coupled to the common cavity mode. The arrangement is described in Fig. \ref{schematic}. 
\begin{figure}[b]
    \centering
    \includegraphics[scale = 0.35]{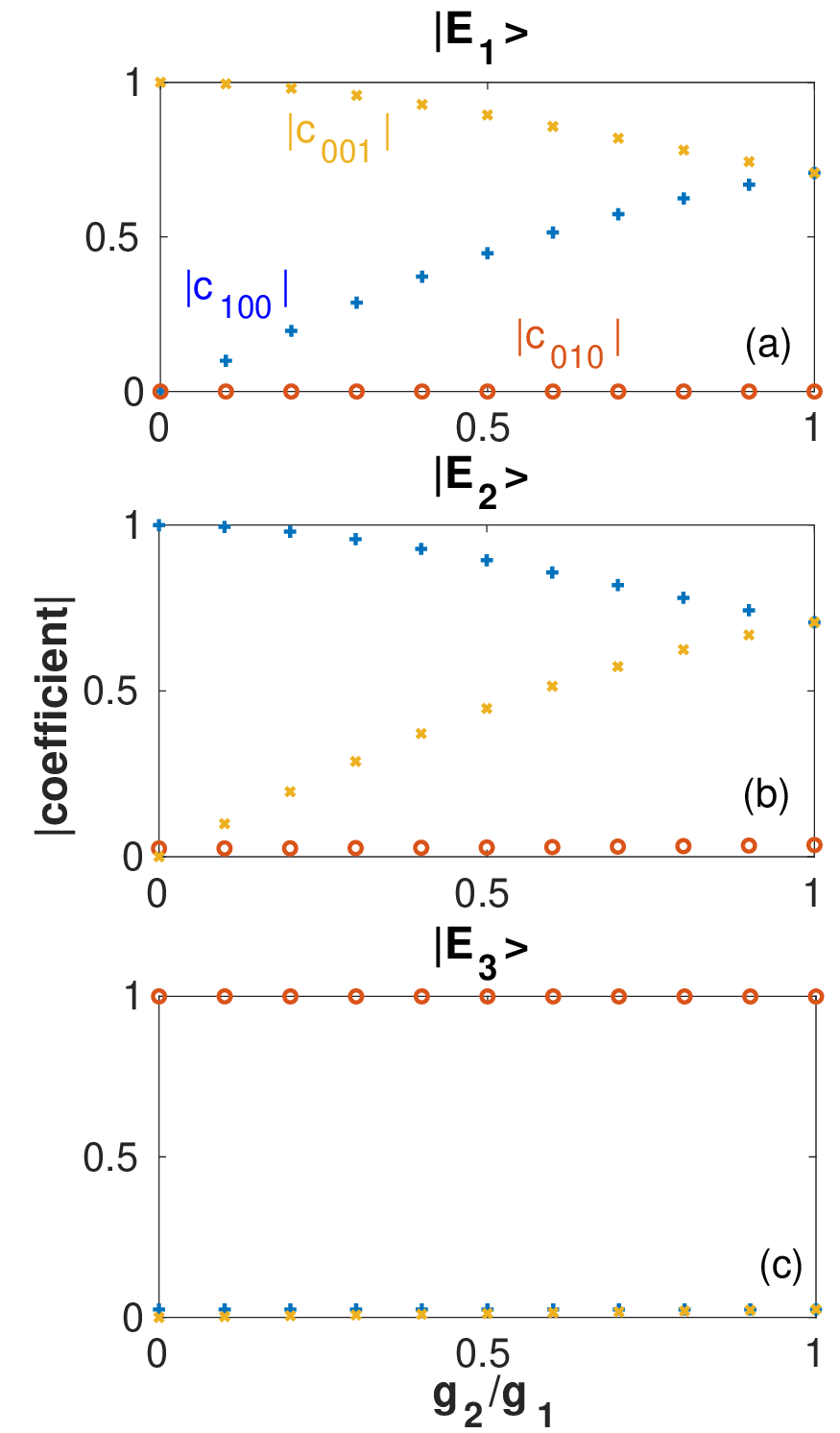}
  %  \captionsetup{justification = centerlast}
    \caption{Absolute values of coefficients corresponding to $|100\rangle$, $|010\rangle$, and $|001\rangle$ for various eigenvectors are plotted with varying $g_2/g_1$. Dispersive case $\omega \neq e_1, e_2$ is considered here. Parameter values used are $\omega=50g_1$, $\epsilon_1=\epsilon_2=10g_1$.}
    \label{fig3}
\end{figure}
\begin{figure}[t]
    \centering
    \includegraphics[scale = 0.32]{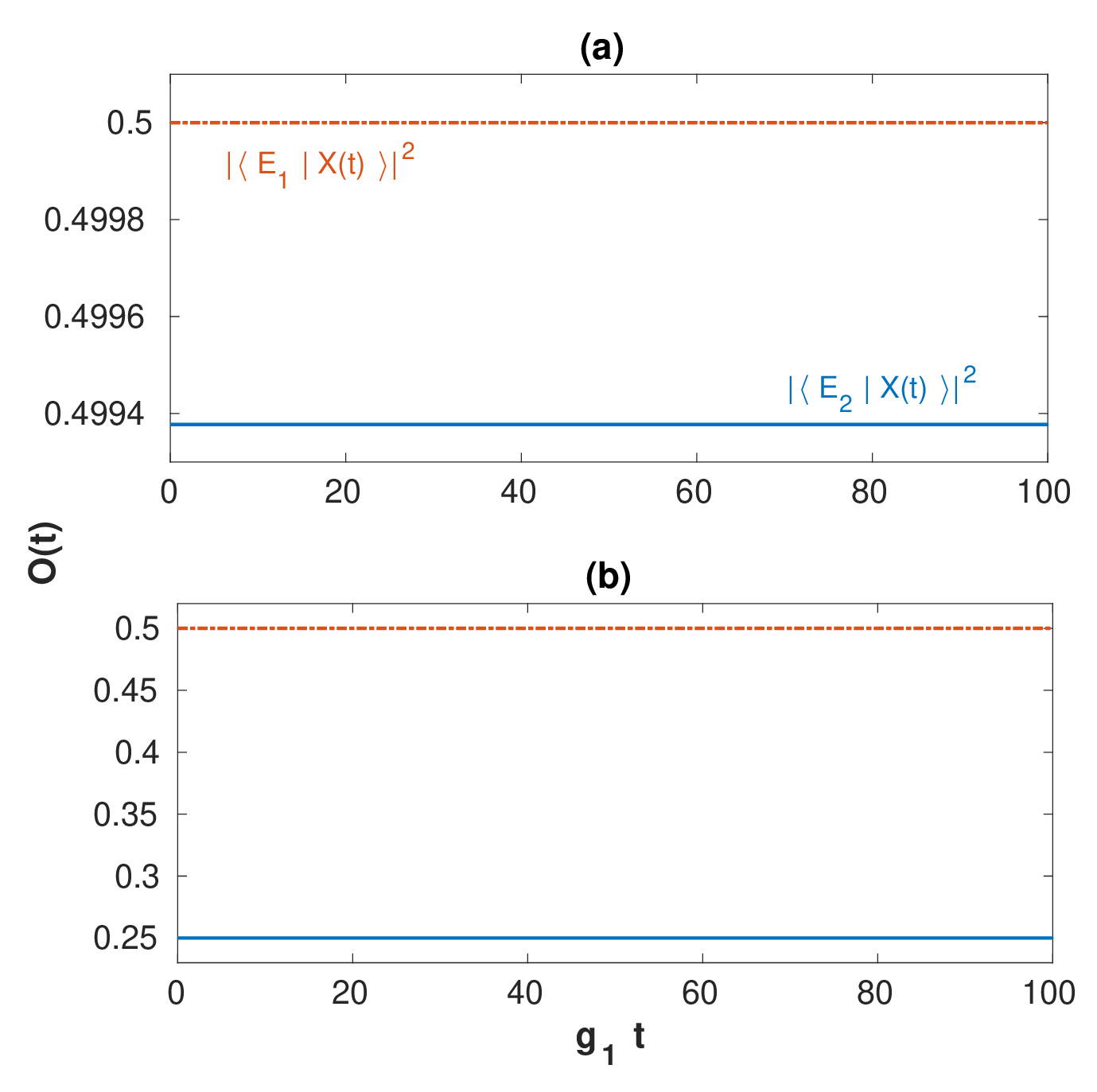}
  %  \captionsetup{justification = centerlast}
    \caption{(a) Time dynamics of the overlap of the evolving state with the eigenstates is plotted when (a) $\epsilon_1=\epsilon_2\neq\omega$ and 
    (b)$\epsilon_1=\epsilon_2=\omega$. Here $g_2/g_1=1$.}
    \label{fig2}
\end{figure}
The objective is to establish quantum correlation between the qubits via the cavity photons. We consider a subspace of single excitation that can be shared between the cavity and qubit degrees of freedom. In this subspace the relevant Fock state basis is $\{|100\rangle, |010\rangle, |001\rangle\}$, where $1$ ($0$) in the first and third position represents excited (ground) state of the first and second qubits, respectively. $0$ ($1$) in the second position represents ground (excited) cavity mode. For simplicity we consider $\epsilon_1=\epsilon_2$ throughout the paper.
\begin{figure}[b]
    \centering
    \includegraphics[scale = 0.35,angle=90, angle=90, angle=90]{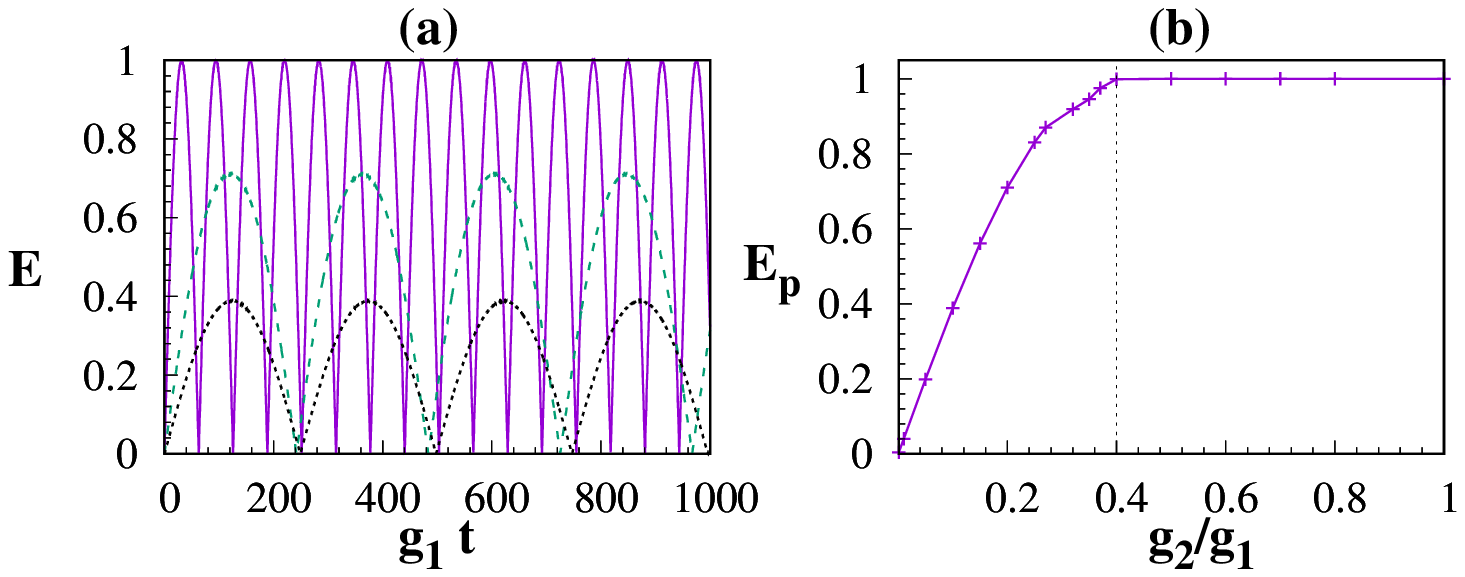}
  %  \captionsetup{justification = centerlast}
    \caption{(a) Dynamics of two-qubit entanglement $E$ is plotted for $g_2/g_1=1$ (solid),$g_2/g_1=0.2$ (dashed), and $g_2/g_1=0.1$ (dotted).    
(b) Peak values of entanglement $E_p$ achieved for various $g_2/g_1$. Vertical dotted line marks the $g_2/g_1$ value where $E_p$ starts deviating from maximum value $1$.}
    \label{fig1}
\end{figure}

\begin{figure}[htb]
    \centering
    \includegraphics[scale = 0.32,angle=90, angle=90, angle=90]{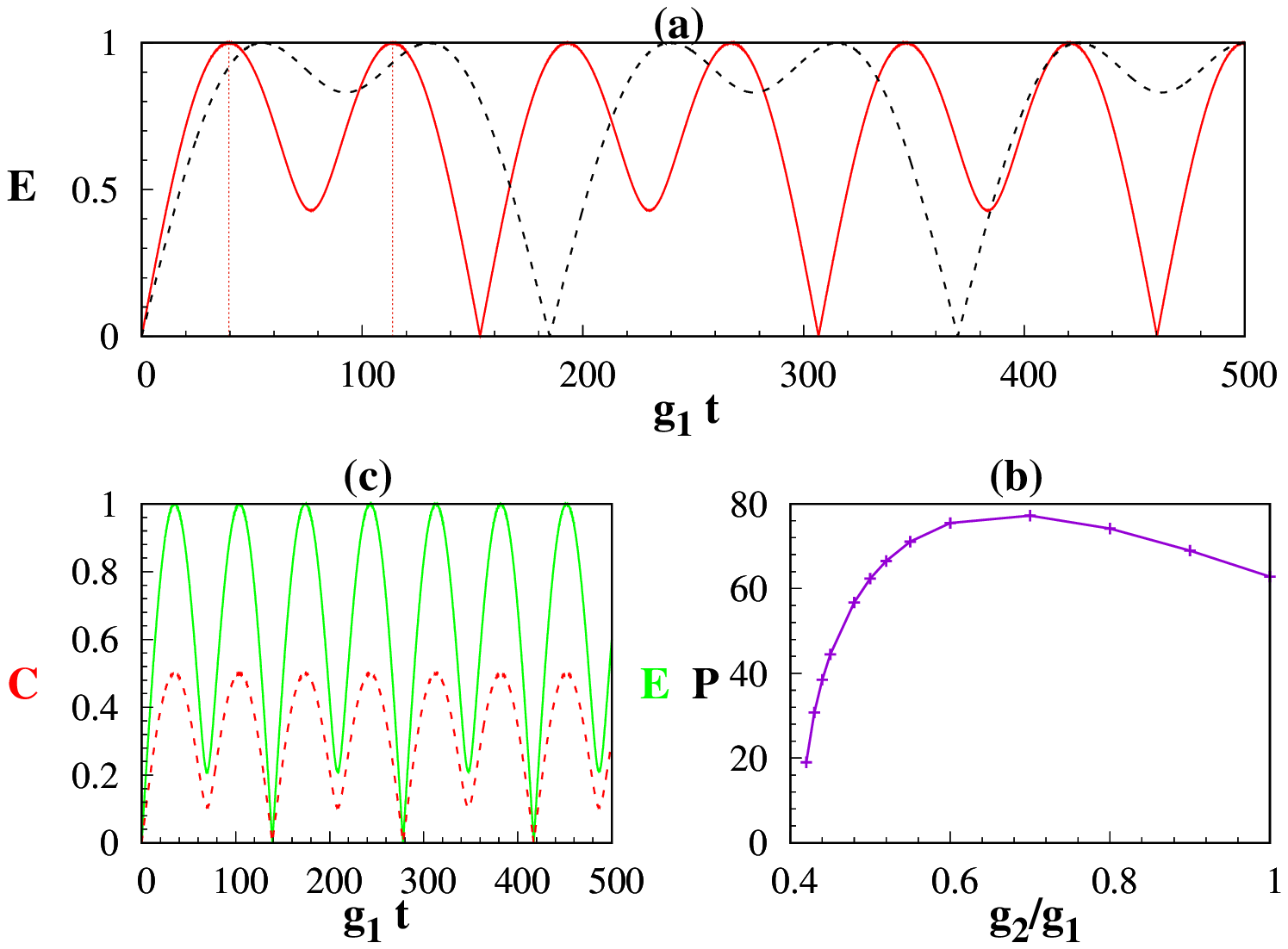}
  %  \captionsetup{justification = centerlast}
    \caption{(a) Entanglement dynamics for $g_2/g_1=0.8$ and $g_2/g_1=0.6$ are presented by solid red and dashed black lines, respectively. 
    The vertical dotted lines mark the times where consecutive MES appear. (b)Minimum time lapse between consecutive appearances of maximally entangled states is plotted against $g_2/g_1$ values. (c) Dynamics of two-qubit coherence C (dashed red) and entanglement is plotted for $g_2/g_1=0.9$.}
    \label{fig4}
\end{figure}
The eigenvalues of the Hamiltonian Eq. \ref{ham} are given by
\begin{eqnarray}
 E_1&=&0\\
 E_2&=&\frac{1}{2}\Big(-e_1+\omega-\sqrt{e_1^2+4(g_1^2+g_2^2)-2e_1\omega+\omega^2}\Big)\nonumber \\
 \\
 E_3&=&\frac{1}{2}\Big(-e_1+\omega+\sqrt{e_1^2+4(g_1^2+g_2^2)-2e_1\omega+\omega^2}\Big)\nonumber \\
\end{eqnarray}
corresponding to the eigenvectors
\begin{eqnarray}
 |E_1\rangle&=& (-\frac{g_2}{g_1})|1 0 0\rangle+|0 0 1\rangle \\
 |E_2\rangle&=& (\frac{g_1}{g_2})|1 0 0\rangle-
 \frac{e_1-\omega+\sqrt{e_1^2+4(g_1^2+g_2^2)-2e_1\omega+\omega^2}}{2g_2} \nonumber \\
 &&~~~~~~~~~~~~~~~~~~~~~~|0 1 0\rangle 
 +|0 0 1 \rangle \nonumber \\
 \\
 |E_3\rangle&=& (\frac{g_1}{g_2})|1 0 0\rangle-
 \frac{e_1-\omega-\sqrt{e_1^2+4(g_1^2+g_2^2)-2e_1\omega+\omega^2}}{2g_2}\nonumber \\
 &&~~~~~~~~~~~~~~~~~~~~~~|0 1 0\rangle 
 +|0 0 1 \rangle\nonumber, \\
\end{eqnarray}
respectively. In the subsequent sections we present the dynamics of the system by initializing the composite system in $|001\rangle$ state, which is a unentangled state having no inter-qubit or qubit-photon correlation. In the next section we will present the numerical results for our model.
 
 \section{Results}\label{result}
Here we present the results for two cases: (i) dispersive case $\omega\neq\epsilon_{1}=\epsilon_2$ and (ii) resonant case $\omega=\epsilon_1=\epsilon_2$. We mainly rely on two types of quantum correlations such as entanglement and coherence. To extract the entanglement we exploit the measure of concurrence. Concurrence $C$ is defined as 
\begin{eqnarray}
 C(\rho)&=& max (0,\lambda_1-\lambda_2-\lambda_3-\lambda_4),
 \label{concurrence}
\end{eqnarray}
where $\rho$ is the two-qubit density matrix and $\lambda_1$, $\lambda_2$, $\lambda_3$, $\lambda_4$ are the eigenvalues of the Hermitian matrix $R$ in decreasing order. $R=\sqrt{\sqrt{\rho} \tilde{\rho} \sqrt{\rho}}$ with $\tilde{\rho}$ being the spin-flipped density matrix defined as
\begin{eqnarray}
\tilde{\rho}&=& (\sigma_y \otimes \sigma_y) \rho^* (\sigma_y \otimes \sigma_y),
\end{eqnarray}
where $\sigma_y$ is Pauli matrix.
\subsection{Dispersive case}
The dispersive setup is based on the fact that there is no real energy exchange between the qubits and the cavity. First, we would like to have an idea about the eigenstates supported by our Hamiltonian given by Eq. \ref{ham}. In Fig. \ref{fig3} we see that eigenstates $|E_1\rangle$ has no projection on the state $|010\rangle$ and $|E_2\rangle$ has negligeable projection on $|010\rangle$. As we initiate our system to the state $|001\rangle$ $|E_1\rangle$ and $|E_2\rangle$ are the relevant eigenstates of the system. We also note that at $g_2/g_1=1$ $|E_1\rangle$ ($|E_2\rangle$) is exactly (almost) a maximally entangled Bell state. As $g_2/g_1$ decreases $|E_1\rangle$ ($|E_2\rangle$) gradually localizes to $|001\rangle$ ($|100\rangle$). Therefore, we make a naive guess of having maximally entangled states (MES) at $g_2/g_1$ values near 1. This is also supported by the observation that, in Fig. \ref{fig2} (a) the overlap of the evolved state $|X(t)\rangle$ with $|E_3\rangle$ is negligeably small. Next, we study the exact dynamics of the model and search for availability of MES at various $g_2/g_1$ values.  In Fig. \ref{fig1} (a) two-qubit entanglement dynamics is plotted for various $g_2/g_1$ values. We observe oscillatory behavior of entanglement $E$ and the fact that the  maximum attainable entanglement ($E_p$) appears less frequently as $g_2/g_1$ decreases. It also reflects that $E_p$ gets reduced with decreasing $g_2/g_1$ value. This fact is well depicted in Fig. \ref{fig1} (b). Here we make an interesting observation that MES ($E_p=1$) is available only up to certain threshold value which is indeed quite far from $g_2/g_1=1$. This result deviates from the inference made by looking at the eigenstates in Fig. \ref{fig3}. We are now interested in the non-uniform coupling $g_2/g_1 \neq 1$ regime where MES is still available. In Fig. \ref{fig4} (a) we have $E$ dynamics for $g_2/g_1=0.8,~0.6$. We see that in every period there is a '$M$'-shaped region consisting two consecutive appearances of MES. The time lapse ($P$) between these two appearances of MES is plotted in Fig. \ref{fig4} (b). Here, $P$ initially increases with $g_2/g_1$ and afterwards gradually decreases. It is important to point out that the $P$ values reaches zero at $g_2/g_1\approx 0.4$ which is the threshold beyond which no MES is available in Fig. \ref{fig1} (b). To explore the dynamical behavior of coherence we plot Fig. \ref{fig4} (c). Here the coherence $C$ is defined as the  absolute value of maximum of the off-diagonal terms present in the two-qubit reduced density matrix. We see qualitatively similar behavior of $C$ and $E$. In the next subsection we analytically treat the problem and give detailed explanation for our numerical observations.    
\begin{figure}[t]
    \centering
    \includegraphics[scale = 0.35]{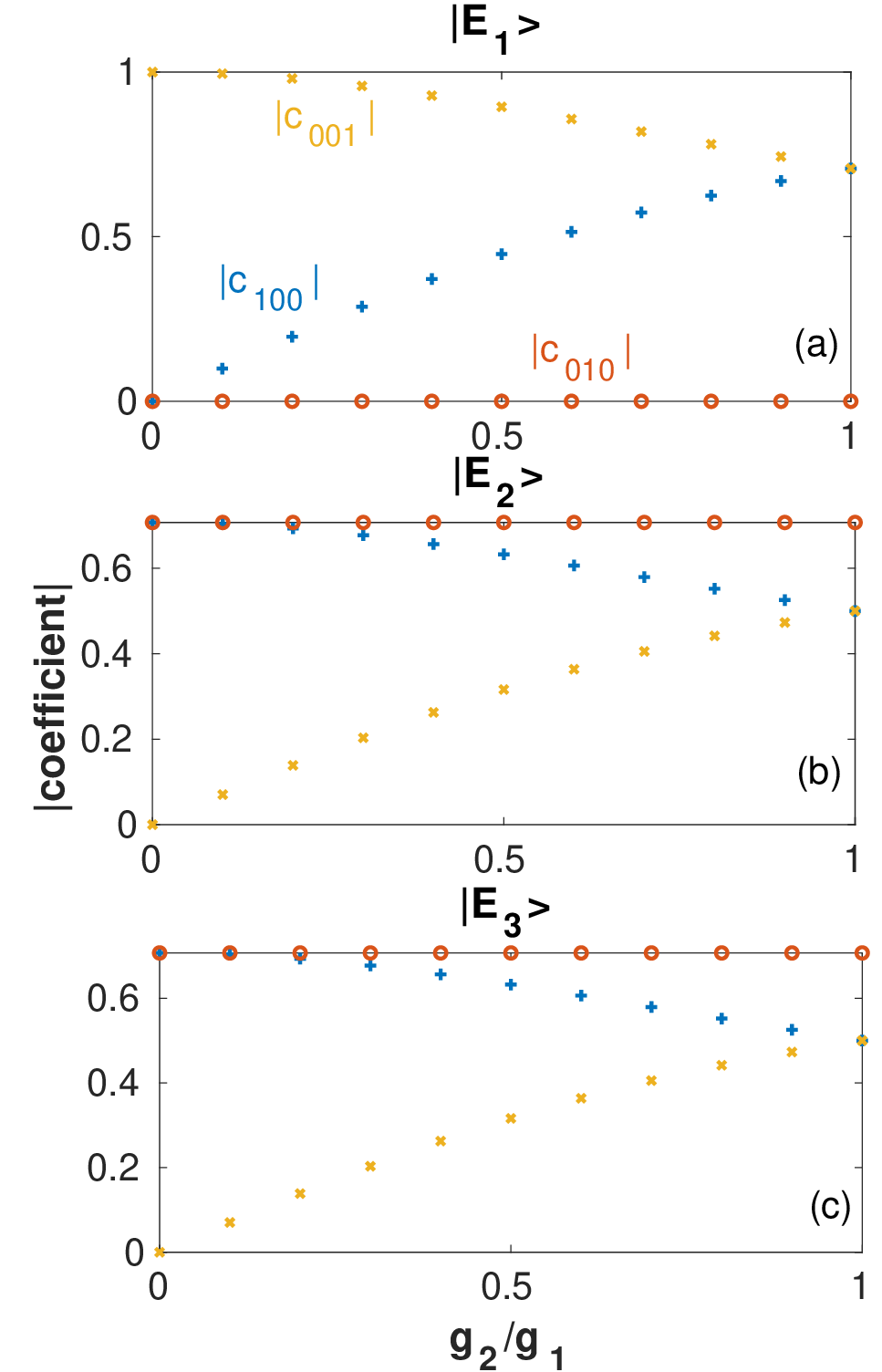}
  %  \captionsetup{justification = centerlast}
    \caption{Absolute values of coefficients corresponding to $|100\rangle$, $|010\rangle$, and $|001\rangle$ for various eigenvectors are plotted with varying $g_2/g_1$. Resonant case $\omega = e_1=e_2$ is considered here.}
    \label{fig5}
\end{figure}
\begin{figure}[b]
    \centering
    \includegraphics[scale = 0.28]{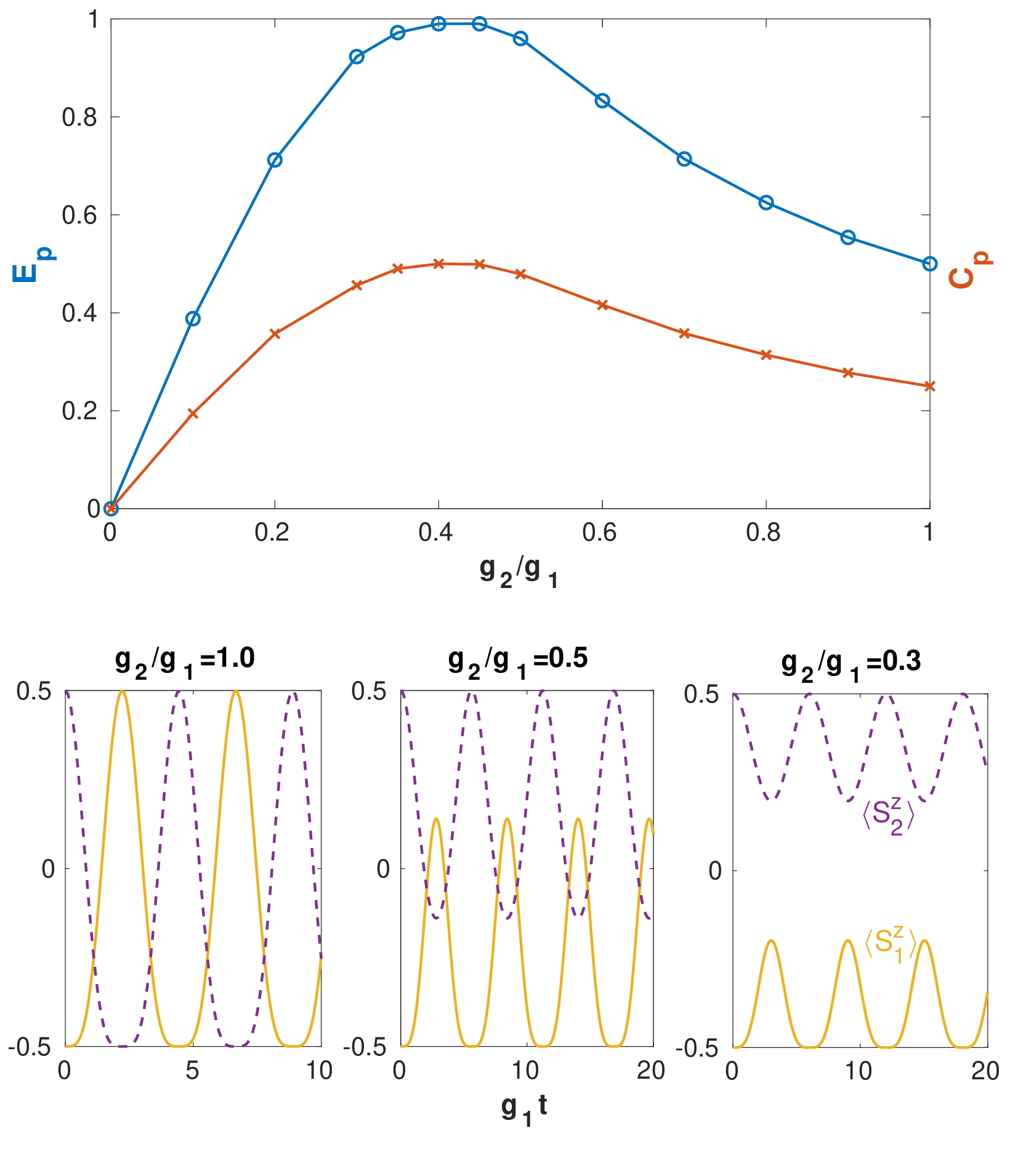}
  %  \captionsetup{justification = centerlast}
    \caption{Top panel:Peak values of 0 ($E_p$) and coherence ($C_p$) are plotted for varying $g_2/g_1$ when $\epsilon_1=\epsilon_2=\omega$. Bottom panel presents dynamics of $\langle s^z_1 \rangle$ and $\langle s^z_2 \rangle$ for different $g_2/g_1$.}
    \label{fig7}
\end{figure}
\subsection{Effective Hamiltonian}
{
Now we would like to calculate the effective Hamiltonian for our Hamiltonian presented in Eq. \ref{ham}. When the off-resonant case is dealt with there is no energy exchange between photonic and qubit degrees of freedom. For $\delta=\epsilon_{i,j}-\omega \gg g_{i,j}$ the relevant transition is between energy conserving states $|100\rangle$ and $|001\rangle$, provided that qubit degree of freedom is excited to begin with. Although no direct interaction between the qubits is available in the Hamiltonian Eq. \ref{ham}, the transition takes place via virtual transition to $|010\rangle$ in second order perturbative process. The rate corresponding to such process is given by
\begin{eqnarray}
 |\frac{\langle 100|H|010\rangle~\langle 010|H|001\rangle}{\delta}|&=&\frac{g_1g_2}{\delta}
\end{eqnarray}
We know that
\begin{eqnarray}
 H_{\rm eff}&=& e^{-\lambda X} H e^{\lambda X} \nonumber \\
 &=& H+\lambda [H,X]+\frac{\lambda^2}{2!}\Big [[H,X],X\Big]+\mathcal{O}(\lambda^3)
\end{eqnarray}
We choose unitary operator such that ${\rm exp}[\lambda X]={\rm exp}[\sum_{i}g_i/\delta(a^{\dagger}_i S_i^{-}-a_i S_i^{+})]$. Retaining terms up to second order in $g_i/\delta$ we get the following effective Hamiltonian.
\begin{eqnarray}
 H_{\rm eff}&=& \omega a^{\dagger}a+\sum^2_{i=1} \epsilon_i S^z_i+\sum^2_{i=1}\frac{g^2_i}{\delta}(a^{\dagger}a+aa^{\dagger})S^z_i\nonumber \\
 &&~~~~~~~~~~~~~~~~+\sum_{i\neq j}\frac{g_ig_j}{2\delta}(S^+_iS^-_j+S^-_iS^+_j).
 \label{effective_ham}
\end{eqnarray}
Third term in Eqn. \ref{effective_ham} represents photon-dependent Stark shift of qubit frequency, whereas, the fourth term is the dipolar inter-qubit 
interaction induced via common photon environment. If the excitation remains with qubit degrees of freedom initially, (e. g., $a^{\dagger}a |001\rangle=0|001\rangle$) the Hamiltonian in Eq. \ref{effective_ham} can be reduced to 
\begin{eqnarray}
 H^{\prime}_{\rm eff}&=&\sum^2_{i=1}\frac{g^2_i}{\delta} S^z_i+\sum_{i\neq j}\frac{g_ig_j}{2\delta}(S^+_iS^-_j+S^-_iS^+_j),
 \label{effective_ham1}
 \end{eqnarray}
 where we consider equal excitations of the qubits, i. e., $\epsilon_1=\epsilon_2$. Now we are interested in obtaining the time evolution of two-qubit states. We primarily obtain the evolution of $|001\rangle$, which is our initial state for all the cases in the manuscript. As we are mainly interested in the two-qubit dynamics and there is no real energy exchange between qubit and photonic degrees of freedom (as can be seen from Eqns. \ref{effective_ham} and \ref{effective_ham1}), we look for the evolution of two-qubit state $|01\rangle_q$ under the Hamiltonian given by Eq. \ref{effective_ham}. Here the first (second) position in $|01\rangle_q$ is reserved for first (second) qubit. One can get
 \begin{eqnarray}
  e^{-iH^{\prime}_{\rm eff}t}|01\rangle_q&=& \{1+\frac{g^2_2}{g^2_1+g^2_2}({\rm exp}[-it(\frac{g^2_1}{\delta}+\frac{g^2_2}{\delta})]-1) \}\nonumber\\
  &&~~~|01\rangle_q\nonumber\\
  &&+\frac{g_1g_2}{g^2_1+g^2_2}({\rm exp}[-it(\frac{g^2_1}{\delta}+\frac{g^2_2}{\delta})]-1) 
  \nonumber\\&&~~~ |10\rangle_q.
  \label{evolved_state}
 \end{eqnarray}
 Let us check the $g_1=g_2$ case, which is represented in Fig. \ref{fig1} (a). The evolved state in Eq. \ref{evolved_state} becomes
 \begin{eqnarray}
 |\psi(t)\rangle=\frac{1}{2}(1+e^{-2ig^2_1t/\delta})|01\rangle_q + \frac{1}{2}(-1+e^{-2ig^2_1t/\delta})|10\rangle_q \nonumber\\
 \end{eqnarray}
For $g^2_1 t/\delta=\pi/4$, $|\psi(t)\rangle$ becomes the maximally entangled EPR state $[|01\rangle_q-i|10\rangle_q]/\sqrt{2}$ up to a global phase factor. The appearance of maximally entangled state occurs at times $t=(2n+1)\pi\delta/(4g^2_1)$ and this is in precise agreement with $g_2/g_1=1$ case in Fig. \ref{fig1} (a).
Now we would like to analyze the $g_1\neq g_2$ cases, especially, when such non-uniform couplings produce maximally entangled states (as presented in Fig. \ref{fig1} (b)). To obtain the condition for maximally entangled state we equate the absolute values of the coefficients of $|01\rangle_q$ and $|10\rangle_q$ in Eq. \ref{evolved_state}. This is because equal-weighted superposition of $|10\rangle_q$ and $|01\rangle_q$ ensures maximally entangled state. We obtain the following condition
\begin{eqnarray}
 {\rm cos} (\frac{g^2_1}{\delta}+\frac{g^2_2}{\delta})t&=& - \frac{(g^2_1-g^2_2)^2}{4g^2_1g^2_2}, \label{cond1}
 \end{eqnarray}
 which in turn implies
 \begin{eqnarray}
 0\le~~~ (g^2_1-g^2_2)^2 &\le& 4g^2_1g^2_2\\
\implies \Big(\frac{g_1}{g_2} \Big)^2+ \Big(\frac{g_2}{g_1} \Big)^2 &\le& 6.
 \label{condition}
\end{eqnarray}
The least value of the left hand side of Eq. \ref{condition} can be 2 (when $g_2/g_1=1$), whereas, the maximum value satisfying the condition is attained when $g_2/g_1 \approx 0.4143$. This feature is clearly reflected in Fig. \ref{fig1} (b), where $E_p=1$ can be attained only till $g_2/g_1\approx 0.4$. We also observe that Eq. \ref{cond1} ensures $\Theta_t=(\frac{g^2_1}{\delta}+\frac{g^2_2}{\delta})t$ to remain in the second and third quadrant  of coordinate system. Therefore, possible values of $\Theta_t$ would be $\pi\pm \theta$, where ${\rm cos} \theta = \frac{(g^2_1-g^2_2)^2}{4g^2_1g^2_2}$ with $0\le \theta \le \pi/2$. However, from the expression of $\Theta_t$ one can easily find out the times at which maximally entangled states are available. Here we refer to Fig. \ref{fig4} (a), where the MES in every period appears at the times same as that obtained from $\Theta_t$. Similarly, the time lapse $P$ in Fig. \ref{fig4} (b) is obtained by exploiting $\Theta_t$.}

\subsection{Resonant case}
In this subsection we delve into the regime of resonant situation when $\omega=\epsilon_1=\epsilon_2$. The immediate consequence of this condition is strong energy exchange between the qubit and the cavity. First, let us look at the eigenstates of the Hamiltonian given by Eq. \ref{ham}. Fig. \ref{fig5} describes that the eigenstates $|E_2\rangle$ and $|E_3\rangle$ are considerably projected on the state $|010\rangle$, whereas, $|E_1\rangle$ has no projection on the photon-excited state $|010\rangle$. Unlike Fig. \ref{fig2} (a), Fig. \ref{fig2} (b) for $g_2/g_1=1$ shows considerable contribution from $|E_2\rangle$ and $|E_3\rangle$ to the evolving state. Consequently, although uniform coupling case is supposed to produce high degree of entanglement (as in the dispersive case), it is actually reduced for resonant case as reflected in Fig. \ref{fig7} top panel. We would like to point out some drastic differences between the outcomes of dispersive and resonant cases. As $g_2/g_1$ reduces beyond a limit, $E_p$ diminishes for the dispersive case. Contrastingly, such behavior for the resonant case appears to be non-monotonic in Fig. \ref{fig7}. The top panel of Fig. \ref{fig7} shows that $E_p$ as well as $C_p$ increase with decreasing $g_2/g_1$ up to some value around $g_2/g_1=0.4$. Beyond this we see a decreasing trend with reducing $g_2/g_1$. To explain this interesting feature we inspect the average values of qubit excitations given by $\langle S^{z}_i\rangle=\langle X(t) | S^z_i | X(t) \rangle$. We know that $\langle S^z_i \rangle$ (for $i=1,2$) with respect to the MES is zero. The $g_2/g_1=1$ case in Fig. \ref{fig7} shows that $\langle S^z_1\rangle$ and $\langle S^z_2\rangle$ intersect each other at a value around $-0.25$, which is quite away from zero value  obtained for MES. We note that, in the beginning of every cycle the rate of excitation gain by the first qubit is slower compared to the rate of excitation released by the second qubit. This lag is due to the fact that part of the excitation is actually being shared with the cavity degrees of freedom. The dominant contribution from $|010\rangle$ resulting reduction of entanglement has already been discussed earlier in this section. Moreover, we also know that the excitation is shared between the qubits via photons. Naturally, it is expected that the enhanced $g_2/g_1$ results in efficient entangling. However, for resonant case this advantage comes in expense of certain excitation shared with the cavity. Therefore, there exists an optimum $g_2/g_1$ that ensures neither an intense inter-qubit exchange nor an enhanced energy sharing with the cavity. This explains the entanglement peak in Fig. \ref{fig7}. The $g_2/g_1=0.5$ figure in the bottom panel of Fig. \ref{fig7} shows that the intersection point of $\langle S^z_1\rangle$ and $\langle S^z_2\rangle$ is much closer to zero. Here the releasing and gaining rates of excitation by the respective qubits are not very different and consequently, energy shared with the cavity is much smaller. Therefore, $E_p$ and $C_p$ for $g_2/g_1=0.5$ are of high quality. Next we focus on the $g_2/g_1=0.3$ case (which is on the other side of the peak in the top panel) in Fig. \ref{fig7}, where $\langle S^z_1 \rangle$ and $\langle S^z_2 \rangle$ curves stay quite away from zero. Here the inter-qubit exchange is so weak that the advantage of small cavity-qubit exchange is not paying off. This is why $E_p$ decreases with decreasing $g_2/g_1$ beyond the peak value. Next we deal with the open-system format of our model.

\section{Open-system dynamics}
Cavities are subjected to photon loss via dissipation channels. Such cavity leakage is one of the obstacles for implementing quantum information processing in cavity-QED systems \cite{PhysRevLett.85.2392,Zhang2014}. While the dispersive case is somewhat immuned to such losses, a resonant case encounters cavity-assisted dissipation resulting short-lived quantumness. In this section we investigate dissipative dynamics of our model and seek a remedy via external drive to the qubit. To attain our purpose we exploit Lindblad master equation given by
\begin{eqnarray}
 \dot{\rho}(t)=-i[H^{\prime},\rho(t)]+\kappa\mathcal{L}[a]+\gamma \sum_{i=1}^2 \mathcal{L}{[\sigma_i^-}],
 \label{lindblad}
\end{eqnarray}
\begin{figure}[htb]
    \centering
    \includegraphics[scale = 0.35,angle=90, angle=90, angle=90]{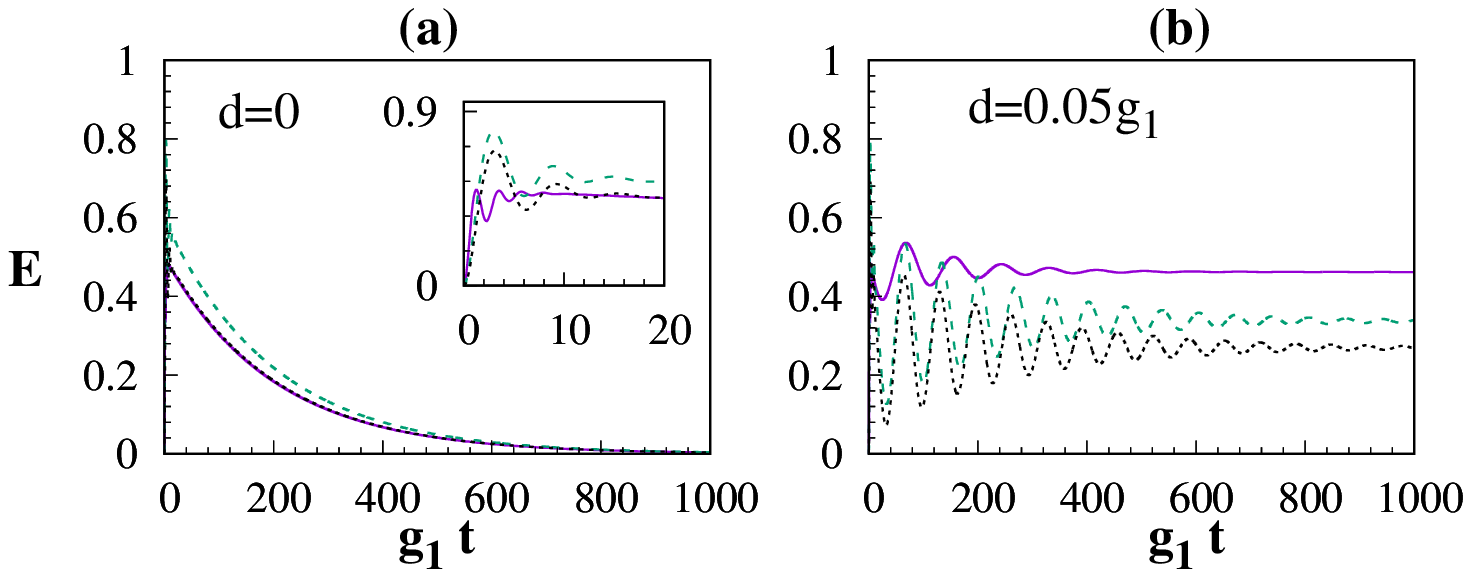}
  %  \captionsetup{justification = centerlast}
    \caption{(a) Dissipative dynamics of two-qubit entanglement $E$ is plotted for $g_2/g_1=1$ (solid),$g_2/g_1=0.4$ (dashed), and $g_2/g_1=0.3$ (dotted). Inset shows short-time dynamics.   
(b) Driven-dissipative dynamics with $d=0.05g_1$ for similar $g_2/g_1$ values as in (a).}
    \label{fig9}
\end{figure}
where $\mathcal{L}[A]=(2A\rho(t)A^{\dagger}-A^{\dagger}A\rho(t)-\rho(t)A^{\dagger}A)/2$. $\kappa$ and $\gamma$ are respectively cavity and qubit dissipation rates. $\rho(t)$ is the reduced density matrix for the two-qubit system. To accommodate qubit drive we introduce the Hamiltonian in the rotating frame of qubit drive frequency and it is written as
\begin{eqnarray}
 H^{\prime}&=&(\omega-\Omega) a^{\dagger}a+\sum_{i=1}^{2} (\epsilon_i-\Omega) S_i+\sum_{i=1}^2 g_i (a^{\dagger} S_i^{-}+{\rm h. c.})\nonumber \\
 && ~~~~~~+d*(S^+_2+S^-_2),
\end{eqnarray}
where $d$ and $\Omega$ are the qubit drive strength and driving field frequency, respectively. Here, we drive just the second qubit to keep the arrangement minimal. The dissipative case is treated by setting $d=0$, $\Omega=0$ making $H^{\prime}$ same as the starting Hamiltonian $H$ in Eq. \ref{ham}, whereas, for driven-dissipative case we set $d\neq0$. Throughout this section we fix $\kappa=g_1$ and $\gamma=0.005g_1$.
In Fig. \ref{fig9} (a) we plot dissipative dynamics of entanglement and observe that the qubits are completely disentangled at long times. However, the $E$ value remains higher for intermediate coupling $g_2/g_1=0.4$ and this is clearly presented in the inset of Fig. \ref{fig9} (a). This observation is in tune with Fig. \ref{fig7} where we found maximum entanglement when $g_2/g_1=0.4$. Now to achieve finite entanglement at long times we employ external qubit drive to the second qubit only. In Fig. \ref{fig9} (b) the driven-dissipative entanglement dynamics saturates to a finite steady state value. We see that the steady-state entanglement increases with increasing coupling $g_2/g_1$. To fine tune our result we plot Fig. \ref{fig10}.
\begin{figure}[htb]
    \centering
    \includegraphics[scale = 0.35,angle=90, angle=90, angle=90]{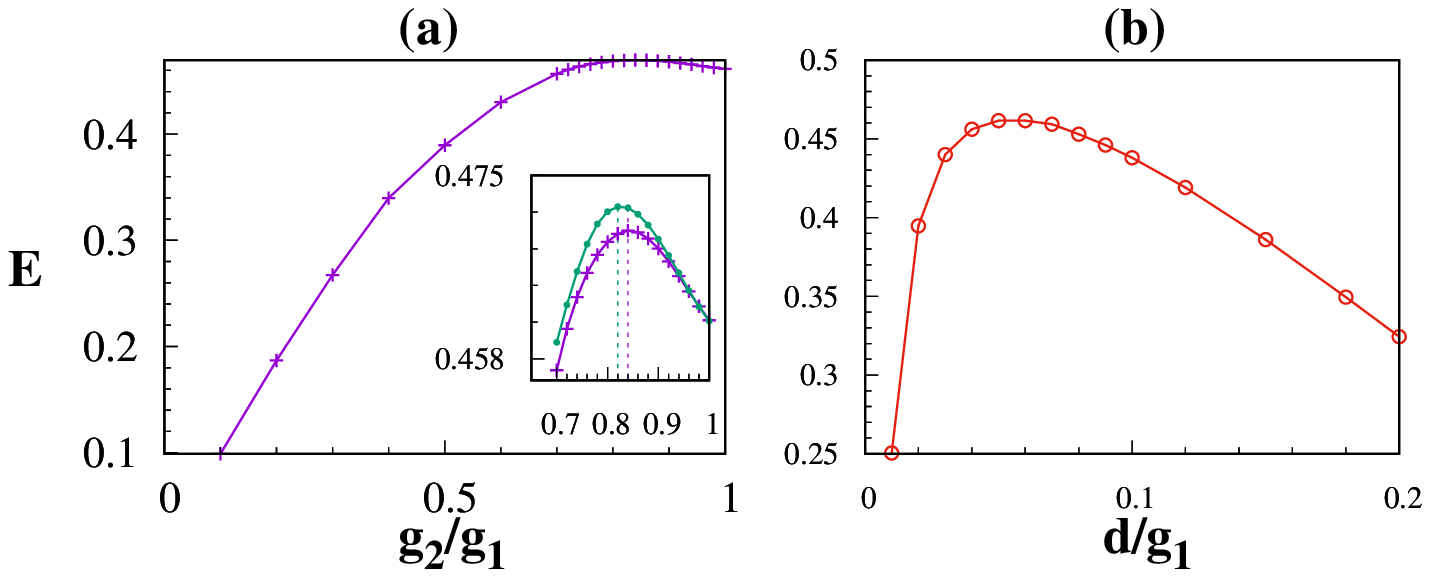}
  %  \captionsetup{justification = centerlast}
    \caption{(a)Steady-state entanglement for varying $g_2/g_1$ when $d=0.05g_1$. Inset presents a portion of the main plot focusing entanglement peaks for $d=.05g_1$ ('+' markers) and $d=.06g_1$ (filled 'o' markers). The peak locations are marked by vertical dashed lines.   
(b) Steady-state entanglement for various $d$ when $g_2/g_1$ is fixed at $1$.}
    \label{fig10}
\end{figure}
In contrast to the apparent monotonicity appearing in Fig. \ref{fig9} (b) the steady-state entanglement in Fig. \ref{fig10} (a) reveals a peak at around $g_2/g_1=0.84$ when $d=0.05g_1$. Similar to the closed-system case in Fig. \ref{fig7} the driven-dissipative case exhibits non-monotonicity in Fig. \ref{fig10} (a). For closed system we have already discussed that the uniform coupling scenario ($g_2/g_1=1$) in resonant case results in two competing effects from the perspective of entanglement generation: (i) enhancement of inter-qubit excitation exchange, which is advantageous for entanglement generation, (ii) elevated energy exchange between qubit and cavity (which is detrimental for inter-qubit entanglement). This results in availability of MES at intermediate $g_2/g_1$ as can be seen in Fig. \ref{fig7}. The peaks in Fig. \ref{fig10} (a) inset appear at the proximity of $g_2/g_1=1$ because the weak driving (compared to the cavity leakage, i. e., $d/\kappa \ll 1$) is unable to populate the cavity considerably. Consequently, the advantage of near-uniform coupling pays off. In the inset of Fig. \ref{fig10} (a) we also notice that the peak value increases and shifts towards lower values of $g_2/g_1$ with increasing $d$. This is just because, with greater drive the chances of cavity excitation becomes greater and a lower $g_2/g_1$ prevents cavity occupation via energy transfer from qubit. Driving a qubit not only prevents it from decaying to ground state but also excites the cavity. Therefore it is important to study the behavior of entanglement with drive strength. In Fig. \ref{fig10} (b) we see that the steady-state entanglement varies non-monotonically with drive strength. This in turn limits the role of drive in retaining entanglement. It is clear from the figure that, too much or too little drive disentangles the qubit pair.
\section{conclusion}
We have studied entanglement generation between two qubits via a common cavity environment. The focus of the study is to investigate the impact of qubit asymmetry on the generated entanglement. We consider parameter values well within the realistic range \cite{RevModPhys.93.025005,PhysRevA.88.023849,PhysRevX.4.031043,meher2022review} making our model experimentally realizable. The driven-dissipative case employs qubit drive to a single qubit only. Stabilizing Bell state through dissipation engineering and single qubit driving reduce experimental complexity \cite{shen2014preparation}. We restrict our analysis only for single-excitation subspace. A high-excitation scenario can exhibit interesting phenomena such as entanglement sudden death and revival\cite{sadiek2019manipulating}. However, our investigation is helpful in estimating entanglement in various degrees of qubit asymmetry and drive intensity. This has a prominent scope for extension to systems with larger components such as qubits in separate cavities \cite{yang2013entanglement} and multiple-qubit systems forming multi-partite entanglement \cite{PhysRevLett.87.230404}. 
\bibliography{bibliography.bib}

\begin{thebibliography}{44}%
\makeatletter
\providecommand \@ifxundefined [1]{%
 \@ifx{#1\undefined}
}%
\providecommand \@ifnum [1]{%
 \ifnum #1\expandafter \@firstoftwo
 \else \expandafter \@secondoftwo
 \fi
}%
\providecommand \@ifx [1]{%
 \ifx #1\expandafter \@firstoftwo
 \else \expandafter \@secondoftwo
 \fi
}%
\providecommand \natexlab [1]{#1}%
\providecommand \enquote  [1]{``#1''}%
\providecommand \bibnamefont  [1]{#1}%
\providecommand \bibfnamefont [1]{#1}%
\providecommand \citenamefont [1]{#1}%
\providecommand \href@noop [0]{\@secondoftwo}%
\providecommand \href [0]{\begingroup \@sanitize@url \@href}%
\providecommand \@href[1]{\@@startlink{#1}\@@href}%
\providecommand \@@href[1]{\endgroup#1\@@endlink}%
\providecommand \@sanitize@url [0]{\catcode `\\12\catcode `\$12\catcode
  `\&12\catcode `\#12\catcode `\^12\catcode `\_12\catcode `\%12\relax}%
\providecommand \@@startlink[1]{}%
\providecommand \@@endlink[0]{}%
\providecommand \url  [0]{\begingroup\@sanitize@url \@url }%
\providecommand \@url [1]{\endgroup\@href {#1}{\urlprefix }}%
\providecommand \urlprefix  [0]{URL }%
\providecommand \Eprint [0]{\href }%
\providecommand \doibase [0]{http://dx.doi.org/}%
\providecommand \selectlanguage [0]{\@gobble}%
\providecommand \bibinfo  [0]{\@secondoftwo}%
\providecommand \bibfield  [0]{\@secondoftwo}%
\providecommand \translation [1]{[#1]}%
\providecommand \BibitemOpen [0]{}%
\providecommand \bibitemStop [0]{}%
\providecommand \bibitemNoStop [0]{.\EOS\space}%
\providecommand \EOS [0]{\spacefactor3000\relax}%
\providecommand \BibitemShut  [1]{\csname bibitem#1\endcsname}%
\let\auto@bib@innerbib\@empty
%</preamble>
\bibitem [{\citenamefont {Nielsen}\ and\ \citenamefont
  {Chuang}(2010)}]{Nielsen2010-zt}%
  \BibitemOpen
  \bibfield  {author} {\bibinfo {author} {\bibfnamefont {M.~A.}\ \bibnamefont
  {Nielsen}}\ and\ \bibinfo {author} {\bibfnamefont {I.~L.}\ \bibnamefont
  {Chuang}},\ }\href@noop {} {\emph {\bibinfo {title} {Quantum Computation and
  Quantum Information}}}\ (\bibinfo  {publisher} {Cambridge University Press},\
  \bibinfo {address} {Cambridge, England},\ \bibinfo {year} {2010})\BibitemShut
  {NoStop}%
\bibitem [{\citenamefont {Horodecki}\ \emph {et~al.}(2009)\citenamefont
  {Horodecki}, \citenamefont {Horodecki}, \citenamefont {Horodecki},\ and\
  \citenamefont {Horodecki}}]{RevModPhys.81.865}%
  \BibitemOpen
  \bibfield  {author} {\bibinfo {author} {\bibfnamefont {R.}~\bibnamefont
  {Horodecki}}, \bibinfo {author} {\bibfnamefont {P.}~\bibnamefont
  {Horodecki}}, \bibinfo {author} {\bibfnamefont {M.}~\bibnamefont
  {Horodecki}}, \ and\ \bibinfo {author} {\bibfnamefont {K.}~\bibnamefont
  {Horodecki}},\ }\href {\doibase 10.1103/RevModPhys.81.865} {\bibfield
  {journal} {\bibinfo  {journal} {Rev. Mod. Phys.}\ }\textbf {\bibinfo {volume}
  {81}},\ \bibinfo {pages} {865} (\bibinfo {year} {2009})}\BibitemShut
  {NoStop}%
\bibitem [{\citenamefont {Ekert}(1991)}]{PhysRevLett.67.661}%
  \BibitemOpen
  \bibfield  {author} {\bibinfo {author} {\bibfnamefont {A.~K.}\ \bibnamefont
  {Ekert}},\ }\href {\doibase 10.1103/PhysRevLett.67.661} {\bibfield  {journal}
  {\bibinfo  {journal} {Phys. Rev. Lett.}\ }\textbf {\bibinfo {volume} {67}},\
  \bibinfo {pages} {661} (\bibinfo {year} {1991})}\BibitemShut {NoStop}%
\bibitem [{\citenamefont {Bennett}\ \emph {et~al.}(1993)\citenamefont
  {Bennett}, \citenamefont {Brassard}, \citenamefont {Cr\'epeau}, \citenamefont
  {Jozsa}, \citenamefont {Peres},\ and\ \citenamefont
  {Wootters}}]{PhysRevLett.70.1895}%
  \BibitemOpen
  \bibfield  {author} {\bibinfo {author} {\bibfnamefont {C.~H.}\ \bibnamefont
  {Bennett}}, \bibinfo {author} {\bibfnamefont {G.}~\bibnamefont {Brassard}},
  \bibinfo {author} {\bibfnamefont {C.}~\bibnamefont {Cr\'epeau}}, \bibinfo
  {author} {\bibfnamefont {R.}~\bibnamefont {Jozsa}}, \bibinfo {author}
  {\bibfnamefont {A.}~\bibnamefont {Peres}}, \ and\ \bibinfo {author}
  {\bibfnamefont {W.~K.}\ \bibnamefont {Wootters}},\ }\href {\doibase
  10.1103/PhysRevLett.70.1895} {\bibfield  {journal} {\bibinfo  {journal}
  {Phys. Rev. Lett.}\ }\textbf {\bibinfo {volume} {70}},\ \bibinfo {pages}
  {1895} (\bibinfo {year} {1993})}\BibitemShut {NoStop}%
\bibitem [{\citenamefont {Bell}(1964)}]{PhysicsPhysiqueFizika.1.195}%
  \BibitemOpen
  \bibfield  {author} {\bibinfo {author} {\bibfnamefont {J.~S.}\ \bibnamefont
  {Bell}},\ }\href {\doibase 10.1103/PhysicsPhysiqueFizika.1.195} {\bibfield
  {journal} {\bibinfo  {journal} {Physics Physique Fizika}\ }\textbf {\bibinfo
  {volume} {1}},\ \bibinfo {pages} {195} (\bibinfo {year} {1964})}\BibitemShut
  {NoStop}%
\bibitem [{\citenamefont {Greenberger}\ \emph {et~al.}(1990)\citenamefont
  {Greenberger}, \citenamefont {Horne}, \citenamefont {Shimony},\ and\
  \citenamefont {Zeilinger}}]{Greenberger1990}%
  \BibitemOpen
  \bibfield  {author} {\bibinfo {author} {\bibfnamefont {D.~M.}\ \bibnamefont
  {Greenberger}}, \bibinfo {author} {\bibfnamefont {M.~A.}\ \bibnamefont
  {Horne}}, \bibinfo {author} {\bibfnamefont {A.}~\bibnamefont {Shimony}}, \
  and\ \bibinfo {author} {\bibfnamefont {A.}~\bibnamefont {Zeilinger}},\ }\href
  {\doibase 10.1119/1.16243} {\bibfield  {journal} {\bibinfo  {journal}
  {American Journal of Physics}\ }\textbf {\bibinfo {volume} {58}},\ \bibinfo
  {pages} {1131–1143} (\bibinfo {year} {1990})}\BibitemShut {NoStop}%
\bibitem [{\citenamefont {Bouwmeester}\ \emph {et~al.}(1999)\citenamefont
  {Bouwmeester}, \citenamefont {Pan}, \citenamefont {Daniell}, \citenamefont
  {Weinfurter},\ and\ \citenamefont {Zeilinger}}]{PhysRevLett.82.1345}%
  \BibitemOpen
  \bibfield  {author} {\bibinfo {author} {\bibfnamefont {D.}~\bibnamefont
  {Bouwmeester}}, \bibinfo {author} {\bibfnamefont {J.-W.}\ \bibnamefont
  {Pan}}, \bibinfo {author} {\bibfnamefont {M.}~\bibnamefont {Daniell}},
  \bibinfo {author} {\bibfnamefont {H.}~\bibnamefont {Weinfurter}}, \ and\
  \bibinfo {author} {\bibfnamefont {A.}~\bibnamefont {Zeilinger}},\ }\href
  {\doibase 10.1103/PhysRevLett.82.1345} {\bibfield  {journal} {\bibinfo
  {journal} {Phys. Rev. Lett.}\ }\textbf {\bibinfo {volume} {82}},\ \bibinfo
  {pages} {1345} (\bibinfo {year} {1999})}\BibitemShut {NoStop}%
\bibitem [{\citenamefont {Pan}\ \emph {et~al.}(2000)\citenamefont {Pan},
  \citenamefont {Bouwmeester}, \citenamefont {Daniell}, \citenamefont
  {Weinfurter},\ and\ \citenamefont {Zeilinger}}]{Pan2000}%
  \BibitemOpen
  \bibfield  {author} {\bibinfo {author} {\bibfnamefont {J.-W.}\ \bibnamefont
  {Pan}}, \bibinfo {author} {\bibfnamefont {D.}~\bibnamefont {Bouwmeester}},
  \bibinfo {author} {\bibfnamefont {M.}~\bibnamefont {Daniell}}, \bibinfo
  {author} {\bibfnamefont {H.}~\bibnamefont {Weinfurter}}, \ and\ \bibinfo
  {author} {\bibfnamefont {A.}~\bibnamefont {Zeilinger}},\ }\href {\doibase
  10.1038/35000514} {\bibfield  {journal} {\bibinfo  {journal} {Nature}\
  }\textbf {\bibinfo {volume} {403}},\ \bibinfo {pages} {515–519} (\bibinfo
  {year} {2000})}\BibitemShut {NoStop}%
\bibitem [{\citenamefont {Su}\ \emph {et~al.}(2007)\citenamefont {Su},
  \citenamefont {Tan}, \citenamefont {Jia}, \citenamefont {Zhang},
  \citenamefont {Xie},\ and\ \citenamefont {Peng}}]{PhysRevLett.98.070502}%
  \BibitemOpen
  \bibfield  {author} {\bibinfo {author} {\bibfnamefont {X.}~\bibnamefont
  {Su}}, \bibinfo {author} {\bibfnamefont {A.}~\bibnamefont {Tan}}, \bibinfo
  {author} {\bibfnamefont {X.}~\bibnamefont {Jia}}, \bibinfo {author}
  {\bibfnamefont {J.}~\bibnamefont {Zhang}}, \bibinfo {author} {\bibfnamefont
  {C.}~\bibnamefont {Xie}}, \ and\ \bibinfo {author} {\bibfnamefont
  {K.}~\bibnamefont {Peng}},\ }\href {\doibase 10.1103/PhysRevLett.98.070502}
  {\bibfield  {journal} {\bibinfo  {journal} {Phys. Rev. Lett.}\ }\textbf
  {\bibinfo {volume} {98}},\ \bibinfo {pages} {070502} (\bibinfo {year}
  {2007})}\BibitemShut {NoStop}%
\bibitem [{\citenamefont {Bugu}\ \emph {et~al.}(2020)\citenamefont {Bugu},
  \citenamefont {Ozaydin}, \citenamefont {Ferrus},\ and\ \citenamefont
  {Kodera}}]{Bugu2020}%
  \BibitemOpen
  \bibfield  {author} {\bibinfo {author} {\bibfnamefont {S.}~\bibnamefont
  {Bugu}}, \bibinfo {author} {\bibfnamefont {F.}~\bibnamefont {Ozaydin}},
  \bibinfo {author} {\bibfnamefont {T.}~\bibnamefont {Ferrus}}, \ and\ \bibinfo
  {author} {\bibfnamefont {T.}~\bibnamefont {Kodera}},\ }\href {\doibase
  10.1038/s41598-020-60299-6} {\bibfield  {journal} {\bibinfo  {journal}
  {Scientific Reports}\ }\textbf {\bibinfo {volume} {10}} (\bibinfo {year}
  {2020}),\ 10.1038/s41598-020-60299-6}\BibitemShut {NoStop}%
\bibitem [{\citenamefont {Blais}\ \emph {et~al.}(2021)\citenamefont {Blais},
  \citenamefont {Grimsmo}, \citenamefont {Girvin},\ and\ \citenamefont
  {Wallraff}}]{RevModPhys.93.025005}%
  \BibitemOpen
  \bibfield  {author} {\bibinfo {author} {\bibfnamefont {A.}~\bibnamefont
  {Blais}}, \bibinfo {author} {\bibfnamefont {A.~L.}\ \bibnamefont {Grimsmo}},
  \bibinfo {author} {\bibfnamefont {S.~M.}\ \bibnamefont {Girvin}}, \ and\
  \bibinfo {author} {\bibfnamefont {A.}~\bibnamefont {Wallraff}},\ }\href
  {\doibase 10.1103/RevModPhys.93.025005} {\bibfield  {journal} {\bibinfo
  {journal} {Rev. Mod. Phys.}\ }\textbf {\bibinfo {volume} {93}},\ \bibinfo
  {pages} {025005} (\bibinfo {year} {2021})}\BibitemShut {NoStop}%
\bibitem [{\citenamefont {Hagley}\ \emph {et~al.}(1997)\citenamefont {Hagley},
  \citenamefont {Ma\^{\i}tre}, \citenamefont {Nogues}, \citenamefont
  {Wunderlich}, \citenamefont {Brune}, \citenamefont {Raimond},\ and\
  \citenamefont {Haroche}}]{PhysRevLett.79.1}%
  \BibitemOpen
  \bibfield  {author} {\bibinfo {author} {\bibfnamefont {E.}~\bibnamefont
  {Hagley}}, \bibinfo {author} {\bibfnamefont {X.}~\bibnamefont {Ma\^{\i}tre}},
  \bibinfo {author} {\bibfnamefont {G.}~\bibnamefont {Nogues}}, \bibinfo
  {author} {\bibfnamefont {C.}~\bibnamefont {Wunderlich}}, \bibinfo {author}
  {\bibfnamefont {M.}~\bibnamefont {Brune}}, \bibinfo {author} {\bibfnamefont
  {J.~M.}\ \bibnamefont {Raimond}}, \ and\ \bibinfo {author} {\bibfnamefont
  {S.}~\bibnamefont {Haroche}},\ }\href {\doibase 10.1103/PhysRevLett.79.1}
  {\bibfield  {journal} {\bibinfo  {journal} {Phys. Rev. Lett.}\ }\textbf
  {\bibinfo {volume} {79}},\ \bibinfo {pages} {1} (\bibinfo {year}
  {1997})}\BibitemShut {NoStop}%
\bibitem [{\citenamefont {Rogers}\ \emph {et~al.}(2017)\citenamefont {Rogers},
  \citenamefont {Cummings}, \citenamefont {Pedrotti},\ and\ \citenamefont
  {Rice}}]{PhysRevA.96.052311}%
  \BibitemOpen
  \bibfield  {author} {\bibinfo {author} {\bibfnamefont {R.}~\bibnamefont
  {Rogers}}, \bibinfo {author} {\bibfnamefont {N.}~\bibnamefont {Cummings}},
  \bibinfo {author} {\bibfnamefont {L.~M.}\ \bibnamefont {Pedrotti}}, \ and\
  \bibinfo {author} {\bibfnamefont {P.}~\bibnamefont {Rice}},\ }\href {\doibase
  10.1103/PhysRevA.96.052311} {\bibfield  {journal} {\bibinfo  {journal} {Phys.
  Rev. A}\ }\textbf {\bibinfo {volume} {96}},\ \bibinfo {pages} {052311}
  (\bibinfo {year} {2017})}\BibitemShut {NoStop}%
\bibitem [{\citenamefont {Rosseau}\ \emph {et~al.}(2014)\citenamefont
  {Rosseau}, \citenamefont {Ha},\ and\ \citenamefont
  {Byrnes}}]{PhysRevA.90.052315}%
  \BibitemOpen
  \bibfield  {author} {\bibinfo {author} {\bibfnamefont {D.}~\bibnamefont
  {Rosseau}}, \bibinfo {author} {\bibfnamefont {Q.}~\bibnamefont {Ha}}, \ and\
  \bibinfo {author} {\bibfnamefont {T.}~\bibnamefont {Byrnes}},\ }\href
  {\doibase 10.1103/PhysRevA.90.052315} {\bibfield  {journal} {\bibinfo
  {journal} {Phys. Rev. A}\ }\textbf {\bibinfo {volume} {90}},\ \bibinfo
  {pages} {052315} (\bibinfo {year} {2014})}\BibitemShut {NoStop}%
\bibitem [{\citenamefont {Yang}\ \emph {et~al.}(2004)\citenamefont {Yang},
  \citenamefont {Chu},\ and\ \citenamefont {Han}}]{PhysRevLett.92.117902}%
  \BibitemOpen
  \bibfield  {author} {\bibinfo {author} {\bibfnamefont {C.-P.}\ \bibnamefont
  {Yang}}, \bibinfo {author} {\bibfnamefont {S.-I.}\ \bibnamefont {Chu}}, \
  and\ \bibinfo {author} {\bibfnamefont {S.}~\bibnamefont {Han}},\ }\href
  {\doibase 10.1103/PhysRevLett.92.117902} {\bibfield  {journal} {\bibinfo
  {journal} {Phys. Rev. Lett.}\ }\textbf {\bibinfo {volume} {92}},\ \bibinfo
  {pages} {117902} (\bibinfo {year} {2004})}\BibitemShut {NoStop}%
\bibitem [{\citenamefont {Rauschenbeutel}\ \emph {et~al.}(2000)\citenamefont
  {Rauschenbeutel}, \citenamefont {Nogues}, \citenamefont {Osnaghi},
  \citenamefont {Bertet}, \citenamefont {Brune}, \citenamefont {Raimond},\ and\
  \citenamefont {Haroche}}]{Rauschenbeutel2000}%
  \BibitemOpen
  \bibfield  {author} {\bibinfo {author} {\bibfnamefont {A.}~\bibnamefont
  {Rauschenbeutel}}, \bibinfo {author} {\bibfnamefont {G.}~\bibnamefont
  {Nogues}}, \bibinfo {author} {\bibfnamefont {S.}~\bibnamefont {Osnaghi}},
  \bibinfo {author} {\bibfnamefont {P.}~\bibnamefont {Bertet}}, \bibinfo
  {author} {\bibfnamefont {M.}~\bibnamefont {Brune}}, \bibinfo {author}
  {\bibfnamefont {J.-M.}\ \bibnamefont {Raimond}}, \ and\ \bibinfo {author}
  {\bibfnamefont {S.}~\bibnamefont {Haroche}},\ }\href {\doibase
  10.1126/science.288.5473.2024} {\bibfield  {journal} {\bibinfo  {journal}
  {Science}\ }\textbf {\bibinfo {volume} {288}},\ \bibinfo {pages}
  {2024–2028} (\bibinfo {year} {2000})}\BibitemShut {NoStop}%
\bibitem [{\citenamefont {Zheng}\ and\ \citenamefont
  {Guo}(2000)}]{PhysRevLett.85.2392}%
  \BibitemOpen
  \bibfield  {author} {\bibinfo {author} {\bibfnamefont {S.-B.}\ \bibnamefont
  {Zheng}}\ and\ \bibinfo {author} {\bibfnamefont {G.-C.}\ \bibnamefont
  {Guo}},\ }\href {\doibase 10.1103/PhysRevLett.85.2392} {\bibfield  {journal}
  {\bibinfo  {journal} {Phys. Rev. Lett.}\ }\textbf {\bibinfo {volume} {85}},\
  \bibinfo {pages} {2392} (\bibinfo {year} {2000})}\BibitemShut {NoStop}%
\bibitem [{\citenamefont {Turchette}\ \emph {et~al.}(1995)\citenamefont
  {Turchette}, \citenamefont {Hood}, \citenamefont {Lange}, \citenamefont
  {Mabuchi},\ and\ \citenamefont {Kimble}}]{PhysRevLett.75.4710}%
  \BibitemOpen
  \bibfield  {author} {\bibinfo {author} {\bibfnamefont {Q.~A.}\ \bibnamefont
  {Turchette}}, \bibinfo {author} {\bibfnamefont {C.~J.}\ \bibnamefont {Hood}},
  \bibinfo {author} {\bibfnamefont {W.}~\bibnamefont {Lange}}, \bibinfo
  {author} {\bibfnamefont {H.}~\bibnamefont {Mabuchi}}, \ and\ \bibinfo
  {author} {\bibfnamefont {H.~J.}\ \bibnamefont {Kimble}},\ }\href {\doibase
  10.1103/PhysRevLett.75.4710} {\bibfield  {journal} {\bibinfo  {journal}
  {Phys. Rev. Lett.}\ }\textbf {\bibinfo {volume} {75}},\ \bibinfo {pages}
  {4710} (\bibinfo {year} {1995})}\BibitemShut {NoStop}%
\bibitem [{\citenamefont {Barenco}\ \emph {et~al.}(1995)\citenamefont
  {Barenco}, \citenamefont {Deutsch}, \citenamefont {Ekert},\ and\
  \citenamefont {Jozsa}}]{PhysRevLett.74.4083}%
  \BibitemOpen
  \bibfield  {author} {\bibinfo {author} {\bibfnamefont {A.}~\bibnamefont
  {Barenco}}, \bibinfo {author} {\bibfnamefont {D.}~\bibnamefont {Deutsch}},
  \bibinfo {author} {\bibfnamefont {A.}~\bibnamefont {Ekert}}, \ and\ \bibinfo
  {author} {\bibfnamefont {R.}~\bibnamefont {Jozsa}},\ }\href {\doibase
  10.1103/PhysRevLett.74.4083} {\bibfield  {journal} {\bibinfo  {journal}
  {Phys. Rev. Lett.}\ }\textbf {\bibinfo {volume} {74}},\ \bibinfo {pages}
  {4083} (\bibinfo {year} {1995})}\BibitemShut {NoStop}%
\bibitem [{\citenamefont {Leghtas}\ \emph {et~al.}(2013)\citenamefont
  {Leghtas}, \citenamefont {Vool}, \citenamefont {Shankar}, \citenamefont
  {Hatridge}, \citenamefont {Girvin}, \citenamefont {Devoret},\ and\
  \citenamefont {Mirrahimi}}]{PhysRevA.88.023849}%
  \BibitemOpen
  \bibfield  {author} {\bibinfo {author} {\bibfnamefont {Z.}~\bibnamefont
  {Leghtas}}, \bibinfo {author} {\bibfnamefont {U.}~\bibnamefont {Vool}},
  \bibinfo {author} {\bibfnamefont {S.}~\bibnamefont {Shankar}}, \bibinfo
  {author} {\bibfnamefont {M.}~\bibnamefont {Hatridge}}, \bibinfo {author}
  {\bibfnamefont {S.~M.}\ \bibnamefont {Girvin}}, \bibinfo {author}
  {\bibfnamefont {M.~H.}\ \bibnamefont {Devoret}}, \ and\ \bibinfo {author}
  {\bibfnamefont {M.}~\bibnamefont {Mirrahimi}},\ }\href {\doibase
  10.1103/PhysRevA.88.023849} {\bibfield  {journal} {\bibinfo  {journal} {Phys.
  Rev. A}\ }\textbf {\bibinfo {volume} {88}},\ \bibinfo {pages} {023849}
  (\bibinfo {year} {2013})}\BibitemShut {NoStop}%
\bibitem [{\citenamefont {Kastoryano}\ \emph {et~al.}(2011)\citenamefont
  {Kastoryano}, \citenamefont {Reiter},\ and\ \citenamefont
  {S\o{}rensen}}]{PhysRevLett.106.090502}%
  \BibitemOpen
  \bibfield  {author} {\bibinfo {author} {\bibfnamefont {M.~J.}\ \bibnamefont
  {Kastoryano}}, \bibinfo {author} {\bibfnamefont {F.}~\bibnamefont {Reiter}},
  \ and\ \bibinfo {author} {\bibfnamefont {A.~S.}\ \bibnamefont
  {S\o{}rensen}},\ }\href {\doibase 10.1103/PhysRevLett.106.090502} {\bibfield
  {journal} {\bibinfo  {journal} {Phys. Rev. Lett.}\ }\textbf {\bibinfo
  {volume} {106}},\ \bibinfo {pages} {090502} (\bibinfo {year}
  {2011})}\BibitemShut {NoStop}%
\bibitem [{\citenamefont {Shen}\ \emph {et~al.}(2011)\citenamefont {Shen},
  \citenamefont {Chen}, \citenamefont {Yang}, \citenamefont {Wu},\ and\
  \citenamefont {Zheng}}]{PhysRevA.84.064302}%
  \BibitemOpen
  \bibfield  {author} {\bibinfo {author} {\bibfnamefont {L.-T.}\ \bibnamefont
  {Shen}}, \bibinfo {author} {\bibfnamefont {X.-Y.}\ \bibnamefont {Chen}},
  \bibinfo {author} {\bibfnamefont {Z.-B.}\ \bibnamefont {Yang}}, \bibinfo
  {author} {\bibfnamefont {H.-Z.}\ \bibnamefont {Wu}}, \ and\ \bibinfo {author}
  {\bibfnamefont {S.-B.}\ \bibnamefont {Zheng}},\ }\href {\doibase
  10.1103/PhysRevA.84.064302} {\bibfield  {journal} {\bibinfo  {journal} {Phys.
  Rev. A}\ }\textbf {\bibinfo {volume} {84}},\ \bibinfo {pages} {064302}
  (\bibinfo {year} {2011})}\BibitemShut {NoStop}%
\bibitem [{\citenamefont {Reiter}\ \emph {et~al.}(2013)\citenamefont {Reiter},
  \citenamefont {Tornberg}, \citenamefont {Johansson},\ and\ \citenamefont
  {S\o{}rensen}}]{PhysRevA.88.032317}%
  \BibitemOpen
  \bibfield  {author} {\bibinfo {author} {\bibfnamefont {F.}~\bibnamefont
  {Reiter}}, \bibinfo {author} {\bibfnamefont {L.}~\bibnamefont {Tornberg}},
  \bibinfo {author} {\bibfnamefont {G.}~\bibnamefont {Johansson}}, \ and\
  \bibinfo {author} {\bibfnamefont {A.~S.}\ \bibnamefont {S\o{}rensen}},\
  }\href {\doibase 10.1103/PhysRevA.88.032317} {\bibfield  {journal} {\bibinfo
  {journal} {Phys. Rev. A}\ }\textbf {\bibinfo {volume} {88}},\ \bibinfo
  {pages} {032317} (\bibinfo {year} {2013})}\BibitemShut {NoStop}%
\bibitem [{\citenamefont {Ma\^{\i}tre}\ \emph {et~al.}(1997)\citenamefont
  {Ma\^{\i}tre}, \citenamefont {Hagley}, \citenamefont {Nogues}, \citenamefont
  {Wunderlich}, \citenamefont {Goy}, \citenamefont {Brune}, \citenamefont
  {Raimond},\ and\ \citenamefont {Haroche}}]{PhysRevLett.79.769}%
  \BibitemOpen
  \bibfield  {author} {\bibinfo {author} {\bibfnamefont {X.}~\bibnamefont
  {Ma\^{\i}tre}}, \bibinfo {author} {\bibfnamefont {E.}~\bibnamefont {Hagley}},
  \bibinfo {author} {\bibfnamefont {G.}~\bibnamefont {Nogues}}, \bibinfo
  {author} {\bibfnamefont {C.}~\bibnamefont {Wunderlich}}, \bibinfo {author}
  {\bibfnamefont {P.}~\bibnamefont {Goy}}, \bibinfo {author} {\bibfnamefont
  {M.}~\bibnamefont {Brune}}, \bibinfo {author} {\bibfnamefont {J.~M.}\
  \bibnamefont {Raimond}}, \ and\ \bibinfo {author} {\bibfnamefont
  {S.}~\bibnamefont {Haroche}},\ }\href {\doibase 10.1103/PhysRevLett.79.769}
  {\bibfield  {journal} {\bibinfo  {journal} {Phys. Rev. Lett.}\ }\textbf
  {\bibinfo {volume} {79}},\ \bibinfo {pages} {769} (\bibinfo {year}
  {1997})}\BibitemShut {NoStop}%
\bibitem [{\citenamefont {Zheng}(2001)}]{PhysRevLett.87.230404}%
  \BibitemOpen
  \bibfield  {author} {\bibinfo {author} {\bibfnamefont {S.-B.}\ \bibnamefont
  {Zheng}},\ }\href {\doibase 10.1103/PhysRevLett.87.230404} {\bibfield
  {journal} {\bibinfo  {journal} {Phys. Rev. Lett.}\ }\textbf {\bibinfo
  {volume} {87}},\ \bibinfo {pages} {230404} (\bibinfo {year}
  {2001})}\BibitemShut {NoStop}%
\bibitem [{\citenamefont {Zhang}\ \emph {et~al.}(2014)\citenamefont {Zhang},
  \citenamefont {Wang}, \citenamefont {Liu},\ and\ \citenamefont
  {Liu}}]{Zhang2014}%
  \BibitemOpen
  \bibfield  {author} {\bibinfo {author} {\bibfnamefont {F.}~\bibnamefont
  {Zhang}}, \bibinfo {author} {\bibfnamefont {D.}~\bibnamefont {Wang}},
  \bibinfo {author} {\bibfnamefont {K.}~\bibnamefont {Liu}}, \ and\ \bibinfo
  {author} {\bibfnamefont {C.}~\bibnamefont {Liu}},\ }\href {\doibase
  10.1007/s10773-014-2446-5} {\bibfield  {journal} {\bibinfo  {journal}
  {International Journal of Theoretical Physics}\ }\textbf {\bibinfo {volume}
  {54}},\ \bibinfo {pages} {2258–2260} (\bibinfo {year} {2014})}\BibitemShut
  {NoStop}%
\bibitem [{\citenamefont {Majer}\ \emph {et~al.}(2007)\citenamefont {Majer},
  \citenamefont {Chow}, \citenamefont {Gambetta}, \citenamefont {Koch},
  \citenamefont {Johnson}, \citenamefont {Schreier}, \citenamefont {Frunzio},
  \citenamefont {Schuster}, \citenamefont {Houck}, \citenamefont {Wallraff},
  \citenamefont {Blais}, \citenamefont {Devoret}, \citenamefont {Girvin},\ and\
  \citenamefont {Schoelkopf}}]{majer2007}%
  \BibitemOpen
  \bibfield  {author} {\bibinfo {author} {\bibfnamefont {J.}~\bibnamefont
  {Majer}}, \bibinfo {author} {\bibfnamefont {J.~M.}\ \bibnamefont {Chow}},
  \bibinfo {author} {\bibfnamefont {J.~M.}\ \bibnamefont {Gambetta}}, \bibinfo
  {author} {\bibfnamefont {J.}~\bibnamefont {Koch}}, \bibinfo {author}
  {\bibfnamefont {B.~R.}\ \bibnamefont {Johnson}}, \bibinfo {author}
  {\bibfnamefont {J.~A.}\ \bibnamefont {Schreier}}, \bibinfo {author}
  {\bibfnamefont {L.}~\bibnamefont {Frunzio}}, \bibinfo {author} {\bibfnamefont
  {D.~I.}\ \bibnamefont {Schuster}}, \bibinfo {author} {\bibfnamefont {A.~A.}\
  \bibnamefont {Houck}}, \bibinfo {author} {\bibfnamefont {A.}~\bibnamefont
  {Wallraff}}, \bibinfo {author} {\bibfnamefont {A.}~\bibnamefont {Blais}},
  \bibinfo {author} {\bibfnamefont {M.~H.}\ \bibnamefont {Devoret}}, \bibinfo
  {author} {\bibfnamefont {S.~M.}\ \bibnamefont {Girvin}}, \ and\ \bibinfo
  {author} {\bibfnamefont {R.~J.}\ \bibnamefont {Schoelkopf}},\ }\href
  {\doibase 10.1038/nature06184} {\bibfield  {journal} {\bibinfo  {journal}
  {Nature}\ }\textbf {\bibinfo {volume} {449}},\ \bibinfo {pages} {443–447}
  (\bibinfo {year} {2007})}\BibitemShut {NoStop}%
\bibitem [{\citenamefont {Blais}\ \emph {et~al.}(2007)\citenamefont {Blais},
  \citenamefont {Gambetta}, \citenamefont {Wallraff}, \citenamefont {Schuster},
  \citenamefont {Girvin}, \citenamefont {Devoret},\ and\ \citenamefont
  {Schoelkopf}}]{PhysRevA.75.032329}%
  \BibitemOpen
  \bibfield  {author} {\bibinfo {author} {\bibfnamefont {A.}~\bibnamefont
  {Blais}}, \bibinfo {author} {\bibfnamefont {J.}~\bibnamefont {Gambetta}},
  \bibinfo {author} {\bibfnamefont {A.}~\bibnamefont {Wallraff}}, \bibinfo
  {author} {\bibfnamefont {D.~I.}\ \bibnamefont {Schuster}}, \bibinfo {author}
  {\bibfnamefont {S.~M.}\ \bibnamefont {Girvin}}, \bibinfo {author}
  {\bibfnamefont {M.~H.}\ \bibnamefont {Devoret}}, \ and\ \bibinfo {author}
  {\bibfnamefont {R.~J.}\ \bibnamefont {Schoelkopf}},\ }\href {\doibase
  10.1103/PhysRevA.75.032329} {\bibfield  {journal} {\bibinfo  {journal} {Phys.
  Rev. A}\ }\textbf {\bibinfo {volume} {75}},\ \bibinfo {pages} {032329}
  (\bibinfo {year} {2007})}\BibitemShut {NoStop}%
\bibitem [{\citenamefont {Borjans}\ \emph {et~al.}(2019)\citenamefont
  {Borjans}, \citenamefont {Croot}, \citenamefont {Mi}, \citenamefont
  {Gullans},\ and\ \citenamefont {Petta}}]{Borjans2019}%
  \BibitemOpen
  \bibfield  {author} {\bibinfo {author} {\bibfnamefont {F.}~\bibnamefont
  {Borjans}}, \bibinfo {author} {\bibfnamefont {X.~G.}\ \bibnamefont {Croot}},
  \bibinfo {author} {\bibfnamefont {X.}~\bibnamefont {Mi}}, \bibinfo {author}
  {\bibfnamefont {M.~J.}\ \bibnamefont {Gullans}}, \ and\ \bibinfo {author}
  {\bibfnamefont {J.~R.}\ \bibnamefont {Petta}},\ }\href {\doibase
  10.1038/s41586-019-1867-y} {\bibfield  {journal} {\bibinfo  {journal}
  {Nature}\ }\textbf {\bibinfo {volume} {577}},\ \bibinfo {pages} {195–198}
  (\bibinfo {year} {2019})}\BibitemShut {NoStop}%
\bibitem [{\citenamefont {Ritter}\ \emph {et~al.}(2012)\citenamefont {Ritter},
  \citenamefont {N\"{o}lleke}, \citenamefont {Hahn}, \citenamefont {Reiserer},
  \citenamefont {Neuzner}, \citenamefont {Uphoff}, \citenamefont {M\"{u}cke},
  \citenamefont {Figueroa}, \citenamefont {Bochmann},\ and\ \citenamefont
  {Rempe}}]{Ritter2012}%
  \BibitemOpen
  \bibfield  {author} {\bibinfo {author} {\bibfnamefont {S.}~\bibnamefont
  {Ritter}}, \bibinfo {author} {\bibfnamefont {C.}~\bibnamefont {N\"{o}lleke}},
  \bibinfo {author} {\bibfnamefont {C.}~\bibnamefont {Hahn}}, \bibinfo {author}
  {\bibfnamefont {A.}~\bibnamefont {Reiserer}}, \bibinfo {author}
  {\bibfnamefont {A.}~\bibnamefont {Neuzner}}, \bibinfo {author} {\bibfnamefont
  {M.}~\bibnamefont {Uphoff}}, \bibinfo {author} {\bibfnamefont
  {M.}~\bibnamefont {M\"{u}cke}}, \bibinfo {author} {\bibfnamefont
  {E.}~\bibnamefont {Figueroa}}, \bibinfo {author} {\bibfnamefont
  {J.}~\bibnamefont {Bochmann}}, \ and\ \bibinfo {author} {\bibfnamefont
  {G.}~\bibnamefont {Rempe}},\ }\href {\doibase 10.1038/nature11023} {\bibfield
   {journal} {\bibinfo  {journal} {Nature}\ }\textbf {\bibinfo {volume}
  {484}},\ \bibinfo {pages} {195–200} (\bibinfo {year} {2012})}\BibitemShut
  {NoStop}%
\bibitem [{\citenamefont {Ray}\ \emph {et~al.}(2022)\citenamefont {Ray},
  \citenamefont {Dey},\ and\ \citenamefont {Kulkarni}}]{Ray2022}%
  \BibitemOpen
  \bibfield  {author} {\bibinfo {author} {\bibfnamefont {T.}~\bibnamefont
  {Ray}}, \bibinfo {author} {\bibfnamefont {A.}~\bibnamefont {Dey}}, \ and\
  \bibinfo {author} {\bibfnamefont {M.}~\bibnamefont {Kulkarni}},\ }\href
  {\doibase 10.1103/physreva.106.042610} {\bibfield  {journal} {\bibinfo
  {journal} {Physical Review A}\ }\textbf {\bibinfo {volume} {106}} (\bibinfo
  {year} {2022}),\ 10.1103/physreva.106.042610}\BibitemShut {NoStop}%
\bibitem [{\citenamefont {Song}\ \emph {et~al.}(2019)\citenamefont {Song},
  \citenamefont {Xu}, \citenamefont {Li}, \citenamefont {Zhang}, \citenamefont
  {Zhang}, \citenamefont {Liu}, \citenamefont {Guo}, \citenamefont {Wang},
  \citenamefont {Ren}, \citenamefont {Hao}, \citenamefont {Feng}, \citenamefont
  {Fan}, \citenamefont {Zheng}, \citenamefont {Wang}, \citenamefont {Wang},\
  and\ \citenamefont {Zhu}}]{Song2019}%
  \BibitemOpen
  \bibfield  {author} {\bibinfo {author} {\bibfnamefont {C.}~\bibnamefont
  {Song}}, \bibinfo {author} {\bibfnamefont {K.}~\bibnamefont {Xu}}, \bibinfo
  {author} {\bibfnamefont {H.}~\bibnamefont {Li}}, \bibinfo {author}
  {\bibfnamefont {Y.-R.}\ \bibnamefont {Zhang}}, \bibinfo {author}
  {\bibfnamefont {X.}~\bibnamefont {Zhang}}, \bibinfo {author} {\bibfnamefont
  {W.}~\bibnamefont {Liu}}, \bibinfo {author} {\bibfnamefont {Q.}~\bibnamefont
  {Guo}}, \bibinfo {author} {\bibfnamefont {Z.}~\bibnamefont {Wang}}, \bibinfo
  {author} {\bibfnamefont {W.}~\bibnamefont {Ren}}, \bibinfo {author}
  {\bibfnamefont {J.}~\bibnamefont {Hao}}, \bibinfo {author} {\bibfnamefont
  {H.}~\bibnamefont {Feng}}, \bibinfo {author} {\bibfnamefont {H.}~\bibnamefont
  {Fan}}, \bibinfo {author} {\bibfnamefont {D.}~\bibnamefont {Zheng}}, \bibinfo
  {author} {\bibfnamefont {D.-W.}\ \bibnamefont {Wang}}, \bibinfo {author}
  {\bibfnamefont {H.}~\bibnamefont {Wang}}, \ and\ \bibinfo {author}
  {\bibfnamefont {S.-Y.}\ \bibnamefont {Zhu}},\ }\href {\doibase
  10.1126/science.aay0600} {\bibfield  {journal} {\bibinfo  {journal}
  {Science}\ }\textbf {\bibinfo {volume} {365}},\ \bibinfo {pages} {574–577}
  (\bibinfo {year} {2019})}\BibitemShut {NoStop}%
\bibitem [{\citenamefont {Xu}\ \emph {et~al.}(2020)\citenamefont {Xu},
  \citenamefont {Sun}, \citenamefont {Liu}, \citenamefont {Zhang},
  \citenamefont {Li}, \citenamefont {Dong}, \citenamefont {Ren}, \citenamefont
  {Zhang}, \citenamefont {Nori}, \citenamefont {Zheng}, \citenamefont {Fan},\
  and\ \citenamefont {Wang}}]{Xu2020}%
  \BibitemOpen
  \bibfield  {author} {\bibinfo {author} {\bibfnamefont {K.}~\bibnamefont
  {Xu}}, \bibinfo {author} {\bibfnamefont {Z.-H.}\ \bibnamefont {Sun}},
  \bibinfo {author} {\bibfnamefont {W.}~\bibnamefont {Liu}}, \bibinfo {author}
  {\bibfnamefont {Y.-R.}\ \bibnamefont {Zhang}}, \bibinfo {author}
  {\bibfnamefont {H.}~\bibnamefont {Li}}, \bibinfo {author} {\bibfnamefont
  {H.}~\bibnamefont {Dong}}, \bibinfo {author} {\bibfnamefont {W.}~\bibnamefont
  {Ren}}, \bibinfo {author} {\bibfnamefont {P.}~\bibnamefont {Zhang}}, \bibinfo
  {author} {\bibfnamefont {F.}~\bibnamefont {Nori}}, \bibinfo {author}
  {\bibfnamefont {D.}~\bibnamefont {Zheng}}, \bibinfo {author} {\bibfnamefont
  {H.}~\bibnamefont {Fan}}, \ and\ \bibinfo {author} {\bibfnamefont
  {H.}~\bibnamefont {Wang}},\ }\href {\doibase 10.1126/sciadv.aba4935}
  {\bibfield  {journal} {\bibinfo  {journal} {Science Advances}\ }\textbf
  {\bibinfo {volume} {6}} (\bibinfo {year} {2020}),\
  10.1126/sciadv.aba4935}\BibitemShut {NoStop}%
\bibitem [{\citenamefont {Kollár}\ \emph {et~al.}(2019)\citenamefont
  {Kollár}, \citenamefont {Fitzpatrick},\ and\ \citenamefont
  {Houck}}]{Kollr2019}%
  \BibitemOpen
  \bibfield  {author} {\bibinfo {author} {\bibfnamefont {A.~J.}\ \bibnamefont
  {Kollár}}, \bibinfo {author} {\bibfnamefont {M.}~\bibnamefont
  {Fitzpatrick}}, \ and\ \bibinfo {author} {\bibfnamefont {A.~A.}\ \bibnamefont
  {Houck}},\ }\href {\doibase 10.1038/s41586-019-1348-3} {\bibfield  {journal}
  {\bibinfo  {journal} {Nature}\ }\textbf {\bibinfo {volume} {571}},\ \bibinfo
  {pages} {45–50} (\bibinfo {year} {2019})}\BibitemShut {NoStop}%
\bibitem [{\citenamefont {Donaire}\ \emph {et~al.}(2017)\citenamefont
  {Donaire}, \citenamefont {Mu\~noz Casta\~neda},\ and\ \citenamefont
  {Nieto}}]{PhysRevA.96.042714}%
  \BibitemOpen
  \bibfield  {author} {\bibinfo {author} {\bibfnamefont {M.}~\bibnamefont
  {Donaire}}, \bibinfo {author} {\bibfnamefont {J.~M.}\ \bibnamefont {Mu\~noz
  Casta\~neda}}, \ and\ \bibinfo {author} {\bibfnamefont {L.~M.}\ \bibnamefont
  {Nieto}},\ }\href {\doibase 10.1103/PhysRevA.96.042714} {\bibfield  {journal}
  {\bibinfo  {journal} {Phys. Rev. A}\ }\textbf {\bibinfo {volume} {96}},\
  \bibinfo {pages} {042714} (\bibinfo {year} {2017})}\BibitemShut {NoStop}%
\bibitem [{\citenamefont {Singh}\ \emph {et~al.}(2022)\citenamefont {Singh},
  \citenamefont {Anand}, \citenamefont {Pocklington}, \citenamefont {Kemp},\
  and\ \citenamefont {Bernien}}]{PhysRevX.12.011040}%
  \BibitemOpen
  \bibfield  {author} {\bibinfo {author} {\bibfnamefont {K.}~\bibnamefont
  {Singh}}, \bibinfo {author} {\bibfnamefont {S.}~\bibnamefont {Anand}},
  \bibinfo {author} {\bibfnamefont {A.}~\bibnamefont {Pocklington}}, \bibinfo
  {author} {\bibfnamefont {J.~T.}\ \bibnamefont {Kemp}}, \ and\ \bibinfo
  {author} {\bibfnamefont {H.}~\bibnamefont {Bernien}},\ }\href {\doibase
  10.1103/PhysRevX.12.011040} {\bibfield  {journal} {\bibinfo  {journal} {Phys.
  Rev. X}\ }\textbf {\bibinfo {volume} {12}},\ \bibinfo {pages} {011040}
  (\bibinfo {year} {2022})}\BibitemShut {NoStop}%
\bibitem [{\citenamefont {Sheng}\ \emph {et~al.}(2022)\citenamefont {Sheng},
  \citenamefont {Hou}, \citenamefont {He}, \citenamefont {Wang}, \citenamefont
  {Guo}, \citenamefont {Zhuang}, \citenamefont {Mamat}, \citenamefont {Xu},
  \citenamefont {Liu}, \citenamefont {Wang},\ and\ \citenamefont
  {Zhan}}]{PhysRevLett.128.083202}%
  \BibitemOpen
  \bibfield  {author} {\bibinfo {author} {\bibfnamefont {C.}~\bibnamefont
  {Sheng}}, \bibinfo {author} {\bibfnamefont {J.}~\bibnamefont {Hou}}, \bibinfo
  {author} {\bibfnamefont {X.}~\bibnamefont {He}}, \bibinfo {author}
  {\bibfnamefont {K.}~\bibnamefont {Wang}}, \bibinfo {author} {\bibfnamefont
  {R.}~\bibnamefont {Guo}}, \bibinfo {author} {\bibfnamefont {J.}~\bibnamefont
  {Zhuang}}, \bibinfo {author} {\bibfnamefont {B.}~\bibnamefont {Mamat}},
  \bibinfo {author} {\bibfnamefont {P.}~\bibnamefont {Xu}}, \bibinfo {author}
  {\bibfnamefont {M.}~\bibnamefont {Liu}}, \bibinfo {author} {\bibfnamefont
  {J.}~\bibnamefont {Wang}}, \ and\ \bibinfo {author} {\bibfnamefont
  {M.}~\bibnamefont {Zhan}},\ }\href {\doibase 10.1103/PhysRevLett.128.083202}
  {\bibfield  {journal} {\bibinfo  {journal} {Phys. Rev. Lett.}\ }\textbf
  {\bibinfo {volume} {128}},\ \bibinfo {pages} {083202} (\bibinfo {year}
  {2022})}\BibitemShut {NoStop}%
\bibitem [{\citenamefont {Anand}\ \emph {et~al.}(2024)\citenamefont {Anand},
  \citenamefont {Bradley}, \citenamefont {White}, \citenamefont {Ramesh},
  \citenamefont {Singh},\ and\ \citenamefont
  {Bernien}}]{https://doi.org/10.48550/arxiv.2401.10325}%
  \BibitemOpen
  \bibfield  {author} {\bibinfo {author} {\bibfnamefont {S.}~\bibnamefont
  {Anand}}, \bibinfo {author} {\bibfnamefont {C.~E.}\ \bibnamefont {Bradley}},
  \bibinfo {author} {\bibfnamefont {R.}~\bibnamefont {White}}, \bibinfo
  {author} {\bibfnamefont {V.}~\bibnamefont {Ramesh}}, \bibinfo {author}
  {\bibfnamefont {K.}~\bibnamefont {Singh}}, \ and\ \bibinfo {author}
  {\bibfnamefont {H.}~\bibnamefont {Bernien}},\ }\href {\doibase
  10.48550/ARXIV.2401.10325} {\enquote {\bibinfo {title} {A dual-species
  rydberg array},}\ } (\bibinfo {year} {2024})\BibitemShut {NoStop}%
\bibitem [{\citenamefont {Abdelmagid}\ \emph {et~al.}(2023)\citenamefont
  {Abdelmagid}, \citenamefont {Alshehhi},\ and\ \citenamefont
  {Sadiek}}]{Abdelmagid2023}%
  \BibitemOpen
  \bibfield  {author} {\bibinfo {author} {\bibfnamefont {R.}~\bibnamefont
  {Abdelmagid}}, \bibinfo {author} {\bibfnamefont {K.}~\bibnamefont
  {Alshehhi}}, \ and\ \bibinfo {author} {\bibfnamefont {G.}~\bibnamefont
  {Sadiek}},\ }\href {\doibase 10.3390/e25101458} {\bibfield  {journal}
  {\bibinfo  {journal} {Entropy}\ }\textbf {\bibinfo {volume} {25}},\ \bibinfo
  {pages} {1458} (\bibinfo {year} {2023})}\BibitemShut {NoStop}%
\bibitem [{\citenamefont {Raftery}\ \emph {et~al.}(2014)\citenamefont
  {Raftery}, \citenamefont {Sadri}, \citenamefont {Schmidt}, \citenamefont
  {T\"ureci},\ and\ \citenamefont {Houck}}]{PhysRevX.4.031043}%
  \BibitemOpen
  \bibfield  {author} {\bibinfo {author} {\bibfnamefont {J.}~\bibnamefont
  {Raftery}}, \bibinfo {author} {\bibfnamefont {D.}~\bibnamefont {Sadri}},
  \bibinfo {author} {\bibfnamefont {S.}~\bibnamefont {Schmidt}}, \bibinfo
  {author} {\bibfnamefont {H.~E.}\ \bibnamefont {T\"ureci}}, \ and\ \bibinfo
  {author} {\bibfnamefont {A.~A.}\ \bibnamefont {Houck}},\ }\href {\doibase
  10.1103/PhysRevX.4.031043} {\bibfield  {journal} {\bibinfo  {journal} {Phys.
  Rev. X}\ }\textbf {\bibinfo {volume} {4}},\ \bibinfo {pages} {031043}
  (\bibinfo {year} {2014})}\BibitemShut {NoStop}%
\bibitem [{\citenamefont {Meher}\ and\ \citenamefont
  {Sivakumar}(2022)}]{meher2022review}%
  \BibitemOpen
  \bibfield  {author} {\bibinfo {author} {\bibfnamefont {N.}~\bibnamefont
  {Meher}}\ and\ \bibinfo {author} {\bibfnamefont {S.}~\bibnamefont
  {Sivakumar}},\ }\href@noop {} {\bibfield  {journal} {\bibinfo  {journal} {The
  European Physical Journal Plus}\ }\textbf {\bibinfo {volume} {137}},\
  \bibinfo {pages} {985} (\bibinfo {year} {2022})}\BibitemShut {NoStop}%
\bibitem [{\citenamefont {Shen}\ \emph {et~al.}(2014)\citenamefont {Shen},
  \citenamefont {Chen}, \citenamefont {Yang}, \citenamefont {Wu},\ and\
  \citenamefont {Zheng}}]{shen2014preparation}%
  \BibitemOpen
  \bibfield  {author} {\bibinfo {author} {\bibfnamefont {L.-T.}\ \bibnamefont
  {Shen}}, \bibinfo {author} {\bibfnamefont {R.-X.}\ \bibnamefont {Chen}},
  \bibinfo {author} {\bibfnamefont {Z.-B.}\ \bibnamefont {Yang}}, \bibinfo
  {author} {\bibfnamefont {H.-Z.}\ \bibnamefont {Wu}}, \ and\ \bibinfo {author}
  {\bibfnamefont {S.-B.}\ \bibnamefont {Zheng}},\ }\href@noop {} {\bibfield
  {journal} {\bibinfo  {journal} {Optics Letters}\ }\textbf {\bibinfo {volume}
  {39}},\ \bibinfo {pages} {6046} (\bibinfo {year} {2014})}\BibitemShut
  {NoStop}%
\bibitem [{\citenamefont {Sadiek}\ \emph {et~al.}(2019)\citenamefont {Sadiek},
  \citenamefont {Al-Drees},\ and\ \citenamefont
  {Abdallah}}]{sadiek2019manipulating}%
  \BibitemOpen
  \bibfield  {author} {\bibinfo {author} {\bibfnamefont {G.}~\bibnamefont
  {Sadiek}}, \bibinfo {author} {\bibfnamefont {W.}~\bibnamefont {Al-Drees}}, \
  and\ \bibinfo {author} {\bibfnamefont {M.~S.}\ \bibnamefont {Abdallah}},\
  }\href@noop {} {\bibfield  {journal} {\bibinfo  {journal} {Optics express}\
  }\textbf {\bibinfo {volume} {27}},\ \bibinfo {pages} {33799} (\bibinfo {year}
  {2019})}\BibitemShut {NoStop}%
\bibitem [{\citenamefont {Yang}\ \emph {et~al.}(2013)\citenamefont {Yang},
  \citenamefont {Su},\ and\ \citenamefont {Nori}}]{yang2013entanglement}%
  \BibitemOpen
  \bibfield  {author} {\bibinfo {author} {\bibfnamefont {C.-P.}\ \bibnamefont
  {Yang}}, \bibinfo {author} {\bibfnamefont {Q.-P.}\ \bibnamefont {Su}}, \ and\
  \bibinfo {author} {\bibfnamefont {F.}~\bibnamefont {Nori}},\ }\href@noop {}
  {\bibfield  {journal} {\bibinfo  {journal} {New Journal of Physics}\ }\textbf
  {\bibinfo {volume} {15}},\ \bibinfo {pages} {115003} (\bibinfo {year}
  {2013})}\BibitemShut {NoStop}%
\end{thebibliography}%
\end{document}